# Power and Sample Size Calculations for Cluster Randomized Hybrid Type 2 Effectiveness-Implementation Studies

Melody A. Owen[1], Geoffrey M. Curran[2], Justin D. Smith[3], Yacob Tedla[4], Chao Cheng[1], Donna Spiegelman[1]

[1]Center for Methods in Implementation and Prevention Science, Yale University, New Haven, CT, USA
[2]Department of Psychiatry, University of Arkansas for Medical Sciences, Little Rock, AR, USA
[3]Population Health Sciences, University of Utah School of Medicine, Salt Lake City, UT, USA
[4]Division of Epidemiology, Department of Medicine, Vanderbilt University Medical Center, Nashville, TN, USA

November 11, 2024

**Abstract**
Hybrid studies allow investigators to simultaneously study an intervention effectiveness outcome and an implementation research outcome. In particular, type 2 hybrid studies support research that places equal importance on both outcomes rather than focusing on one and secondarily on the other (i.e., type 1 and type 3 studies). Hybrid 2 studies introduce the statistical issue of multiple testing, complicated by the fact that they are typically also cluster randomized trials. Standard statistical methods do not apply in this scenario. Here, we describe the design methodologies available for validly powering hybrid type 2 studies and producing reliable sample size calculations in a cluster-randomized design with a focus on binary outcomes. Through a literature search, 18 publications were identified that included methods relevant to the design of hybrid 2 studies. Five methods were identified, two of which did not account for clustering but are extended in this article to do so, namely the combined outcomes approach and the single 1-degree of freedom combined test. Procedures for powering hybrid 2 studies using these five methods are described and illustrated using input parameters inspired by a study from the Community Intervention to Reduce CardiovascuLar Disease in Chicago (CIRCL-Chicago) Implementation Research Center. In this illustrative example, the intervention effectiveness outcome was controlled blood pressure, and the implementation outcome was reach. The conjunctive test resulted in higher power than the popular p-value adjustment methods, and the newly extended combined outcomes and single 1-DF test were found to be the most powerful among all of the tests.

## 1. Introduction

### 1.1 Background

Introduced in 2012,[1] effectiveness-implementation hybrid studies were offered as a typology to describe ways of blending elements of clinical effectiveness and implementation research into the same study. Three hybrid types were described as a heuristic across a continuum from "mostly intervention effectiveness trial" (type 1) to "mostly implementation trial" (type 3), with a category in the middle which consisted of more equal or balanced evaluations of the effectiveness of an intervention while also evaluating the performance of an implementation strategy or strategy bundle designed to support implementation of the clinical intervention being evaluated (type 2). This paper focuses on the statistical design of hybrid type 2 studies, newly developing methods and comparing them to those already existing. While recent updates and extensions to the hybrid study concept have been published,[2-5] no previous manuscript has focused on the statistical design issues that arise in type 2 studies. Before we develop the formal methodology, some concepts and terms associated with hybrid studies need to be defined.

Clear discussion of hybrid studies requires consistency in terminology. In this manuscript, we use the term "intervention" to refer to the programs, practices, principles, procedures, products, pills, and policies (the "seven Ps")[6] that directly impact participant (i.e., patient, client) health outcomes. In the original hybrid manuscript in 2012, we discussed only "clinical" interventions, but in recent updates on hybrids, we acknowledged the wider range of interventions being studied using the typology. We use the term "implementation strategy" to refer to the support activities/tools used to deliver an intervention. The widely referenced ERIC Study taxonomy of implementation strategies[7,8] describes 73 discrete strategies—e.g., education materials, clinician reminders, etc. It is common in implementation research for bundles of strategies to be studied and compared. We use the term "effectiveness" to refer to intervention outcomes (i.e., symptoms or behavior) and the term "impact" to reference observable effects of implementation outcomes (i.e., adoption of or fidelity to an intervention). A range of implementation outcomes are described later in more detail. While it is most common in hybrid type 1 studies to "add on" evaluations of implementation determinants, processes, and outcomes to individual-level randomized controlled trials, hybrid type 2 and type 3 studies tend to use cluster-randomized trial designs.[4]

While the focus of this paper is on hybrid type 2 studies, a grasp of the basic structures and normative approaches of all three types is helpful. On one end of the continuum, hybrid type 1 studies examine intervention effectiveness while gathering information on implementation in what Stetler et al. refer to as an "implementation-focused process evaluation".[9] Many such studies use an implementation determinant framework[10] like CFIR[11] to guide a qualitative or mixed method evaluation of barriers/facilitators to implementation[12] and potential adaptations to the intervention to improve "implementability."[13] On the other end of the spectrum, hybrid type 3 studies examine the impact of implementation strategies on implementation outcomes while gathering information on effectiveness outcomes associated with the intervention being supported for implementation. Additional study efforts are devoted to evaluating the effectiveness of the intervention delivered during the implementation study. It is common for these trials to compare implementation strategies with randomization happening at a "place" level, such as a clinic.[14] The comparisons might be "head-to-head," comparing two or more "experimental" strategies, or one of the comparators might be "implementation as usual." The analysis of intervention effectiveness is conducted as secondary analysis. Increasingly, cost analyses for implementation are being included.[15] In the case of hybrid type 3 studies, the implementation outcomes are clearly primary. Hybrid type 2 studies concern intervention effectiveness while also formally studying the impact of a well-defined implementation strategy or strategies on implementation outcomes. These studies have two co-primary endpoints. Examples of such studies include the Kaiser Implementation Strategy Bundle for Blood Pressure Control, which is being conducted by the Chicago Implementation Research Center.[16,17] This study, which we describe in greater detail later on, aims to evaluate the use of practice facilitation in a Kaiser bundle in order to improve patient care quality. Equal importance is put on assessing the effectiveness and implementation of the bundle.

Since the original conception of hybrids in 2012, two specific "subtypes" of hybrid type 2 studies have been discussed: a) an effectiveness trial of an intervention nested in an implementation trial, and b) an effectiveness trial of an intervention nested in an a single-arm, preliminary evaluation of an implementation strategy. We often referred to subtype "a" as a hybrid type 2 "dual randomized" study—a patient-level randomized clinical trial nested within a site-level randomized implementation trial. These studies are usually not a 2 × 2 factorial design because the implementation strategies are not used to support implementation of the comparison condition for the intervention, although, it is technically possible with a comparative clinical effectiveness trial where the comparator intervention has similar implementation determinants.

Alternatively, we often referred to subtype "b" as a hybrid type 2 "pilot" study—a patient-level randomized trial nested within a pilot study of one implementation. This subtype is conceptually closer to a hybrid type 1 study than the "dual randomized" subtype. The key difference between a hybrid type 1 and type 2 "pilot" study is that in the type 2, the implementation strategy is well-defined and hypothesized to be feasible and impactful "in the real world"[2] and ideally has been based on a prior explicit implementation determinant evaluation. In a hybrid type 1 studies, it is very common for at least some of the implementation strategies used to support delivery of the intervention in the trial to be infeasible in the real world, and for the effectiveness study to be conducted without previous implementation-focused evaluations of implementation determinants of the intervention. Indeed, implementation determinant evaluation conducted *during* the trial is the central implementation feature of the type 1. As observed by a team of researchers currently conducting a scoping review of the published hybrid study literature (authors Geoffrey Curran and J.D. Smith among them), the hybrid type 2 "pilot" subtype has been used more frequently than the "dual randomized" subtype, likely due to cost and logistical challenges of the dual randomized design.[18]

Curran et al. recently offered reflections on 10 years of the application of the study approaches with recommendations for future use.[4] The authors recommended using the terminology "hybrid *studies*" rather than "hybrid *designs*." Use of the term "design" and the explicit focus on trial methodology in the original paper created confusion for users in the field. Indeed, a full range of research designs are and should be applied within each study type. Supporting this recommendation, the article offers guidance on selecting a research design within each hybrid type. Curran et al. stress that research designs and specific hybrid types are not "inextricably coupled, i.e., a specific research design does not automatically indicate a hybrid type and vice versa." While up until this point it has been common for hybrid studies to apply randomized designs, they have also been conducted using observational designs[19] and designs more commonly used in pilot work. As the current manuscript is focused on hybrid type 2 studies, we attempt to acknowledge and address the range of potential research designs which could be applied within a hybrid 2 approach.

## 1.2 Motivation and Overview

Various statistical design methods exist for analyzing and powering hybrid type 2 studies, but as aforementioned, there have been no manuscripts that specifically discuss the exact statistical methods that can be used for hybrid type 2 studies as the research design options are vast. This manuscript introduces established methods for powering studies with two primary outcomes, as well as extends methods used in individually randomized trials with two primary outcomes to the case of cluster randomized trials. Relevant statistical study design methods were identified through a literature search, which are detailed in Section 1.6 . Before introducing the methods, implementation outcomes are defined alongside real-world examples of them in implementation science research. We also describe conjunctive vs. disjunctive power and provide all the hypothesis frameworks possible for a hybrid type 2 study. Lastly, to illustrate how each statistical design method may be used, we introduce a real-world scenario of preliminary input parameters used to power a hybrid type 2 study.

## 1.3 Defining Implementation Outcomes

Because many biostatisticians are not familiar with implementation research, and because most hybrid type 2 studies have an implementation outcome as one of the two co-primary endpoints, we provide an overview here. Implementation outcomes provide insight by quantifying the extent to which an intervention or program are available to and being accessed by the target population. An implementation outcome framework proposed by Proctor et al. is widely used for defining implementation outcomes, and includes acceptability, appropriate, feasibility, adoption, fidelity, penetration, cost, and sustainability.[20] Some of these outcomes, including acceptability, feasibility, and appropriateness, are less commonly evaluated as quantitative endpoints. The RE-AIM model[21] also provides a popular framework for defining implementation outcomes for interventions. This model includes reach, efficacy, adoption, implementation, and maintenance. Within RE-AIM's Implementation domain is adaptation, cost, and fidelity components. All of these outcomes can be defined quantitatively, and can be used as one of the primary outcomes in a hybrid type 2 study. Table 1 summarizes how each outcome in this framework are defined, and provides an example in implementation research of each outcome.

Implementation outcomes are described and summarized here to remind the reader the important difference between traditional intervention outcomes and implementation outcomes. The former aids in determining whether or not the intervention significantly improves health outcomes, whereas the latter aids in determining the extent to which an intervention or implementation strategy bundle can be successfully realized in the real-world.[22]

## 1.4 Statistical Issues for Hybrid Type 2 Studies

Hybrid studies are advantageous as they allow investigators to draw inferences on both effectiveness outcomes and implementation outcomes, but important statistical issues arise when designing them. The first complication is the statistical issue of multiple testing, which is when we have more than one primary outcome. The presence of co-primary outcomes requires careful consideration during the design phase because choices of how the outcomes are incorporated into the statistical design calculations when powering a study will have important design implications. In multiple testing, one must manage the Type I error rate; how the Type I error rate is defined and accounted for in the presence of two primary outcomes will also have important implications on the study design parameters. Furthermore, effectiveness outcomes and implementation outcomes differ in interpretation and measurement type. This means that the specific outcome types and interpretations must be accounted for when selecting the hypothesis testing framework that is chosen to inform the power analysis. Some frameworks discussed in Section 1.5 may be less desirable in this study design, but none are ruled out because they may be fitting for certain research questions and study goals.

Lastly, effectiveness and implementation outcomes that occur in a clustered manner must be assessed accordingly, for example, when the intervention is randomized and administered at the cluster level (e.g. clinics, hospital wards, village), and when background outcome rates differ by cluster. Although several statistical design methods exist for multiple tests, many have not been adapted for clustered studies. Similarly, they have not been explored in the hybrid type 2 scenario. The statistical methods considered in this manuscript apply to clustered settings and highlight current research that accounts for this.

## 1.5 Hypothesis Testing Frameworks in Hybrid Type 2 Studies

In the presence of two outcomes, at least two hypotheses can be tested. We define $Q$ as the number of primary outcomes; in type 2 hybrid studies, $Q = 2$. Then, in a hypothesis testing framework, we define $\alpha$ to be the family-wise, or overall, false positive rate, that is, the probability of one or more false positives among the $Q$ individual tests.[23] Most commonly, $\alpha = 0.05$, providing a 5% probability of rejecting the null hypothesis as defined for the particular study, when this null is true. Statistical power must also be considered. Three types of power arise in this context; the first is disjunctive power, which is the probability of finding at least one true positive. The second is marginal power, which is the probability of rejecting the null when false for each outcome alone, ignoring multiple testing. The third is conjunctive power, which is the probability of rejecting the null when it is false for both outcomes.

We define $\beta_1^*$ as the effect of the first primary outcome, and $\beta_2^*$ as the effect of the second primary outcome. Since there are two primary outcomes, there are four scenarios to consider when constructing a hypothesis test that evaluates the effect of the intervention, namely: 1) $\beta_1^* = 0$ and $\beta_2^* = 0$, 2) $\beta_1^* = 0$ and $\beta_2^* \neq 0$, 3) $\beta_1^* \neq 0$ and $\beta_2^* = 0$, and 4) $\beta_1^* \neq 0$ and $\beta_2^* \neq 0$. Depending on the research question at hand, these options can be grouped in different ways to create the desired hypothesis test for a hybrid type 2 design.

When interested in disjunctive power, the null hypothesis is $H_0: \beta_1^* = 0$ and $\beta_2^* = 0$ versus the alternative hypothesis, $H_A: \beta_1^* \neq 0$ or $\beta_2^* \neq 0$. This test groups the last three possibilities outlined above together under the alternative hypothesis. For this testing framework, under the alternative hypothesis, the intervention must have an effect on at least one outcome but not necessarily both. Studies with this hypothesis testing framework use disjunctive power because interest is in the evidence in the data for at least one true positive.[24] It is important to note that studies using disjunctive power may reject the null hypothesis not only if the intervention has an effect on exactly one of the outcomes, but also if it has an effect on both as well.

The conjunctive, or intersection-union (IU) test, simply referred to as the conjunctive test, assesses the null hypothesis, $H_0: \beta_1^* = 0$ or $\beta_2^* = 0$, versus the alternative hypothesis, $H_A: \beta_1^* \neq 0$ and $\beta_2^* \neq 0$. Here, the first three scenarios are grouped together and included in the null hypothesis, and the alternative hypothesis requires that the intervention have an effect on both the clinical effectiveness outcome and implementation outcome. This typically results in a more conservative test.[23] Section 3.1 2.5 provides more details on the conjunctive intersection-union test.

Section 2 describes the five key statistical design methods we have identified for handling multiple primary outcomes: p-value adjustments, combined outcomes approach, single 1-degree of freedom (DF) combined test for multiple outcomes, disjunctive 2-DF test for multiple outcomes, and the conjunctive 2-DF test. These are all relevant and viable options for the design of a hybrid type 2 studies. In what follows below, two classical methods, the 2-DF combined test and the combined outcomes approach, is extended to the setting of a hybrid type 2 study with clustered data. The advantages and disadvantages of each method is discussed, and an illustrative example that uses these methods is provided to exemplify their use.

### 1.6 Literature Search for Statistical Design Methods

To ensure we were aware of all relevant recent publications, we searched for such in Google Scholar and PubMed. The search was conducted in February 2022 and updated in March 2024. A total of 18 publications were identified that were relevant to study design methodology for hybrid type 2 studies, and from these publications, five key study design methods were identified. Search terms for the literature search were: Co-primary endpoints, cluster randomized trial study design, multiple endpoints, multiple outcomes, study design for multiple outcomes, implementation and clinical health outcomes, hybrid studies, hybrid design methodology.

### 1.7 Notation for Cluster Randomized Trials (CRTs)

Before introducing the cluster randomized study that is used for our illustrative example, we first define the notation for CRTs. Let $Q = 2$ be the number of primary outcomes with outcome index $q = 1, 2$. There are two treatment groups, denoted $i = 1, 2$. Following notation from Donner and Klar, $r$ is the ratio of untreated to treated clusters, which is expressed as $K_2^{(q)} = rK_1^{(q)}$.[25] Under equal treatment allocation, we set the number of clusters in each treatment group to be the same, where $r = 1$ and $K_1^{(q)} = K_2^{(q)} = K^{(q)}$. Note that if we are ignoring multiple comparisons and focusing on one outcome, the number of clusters in each treatment group under equal allocation would simply be denoted as $K$. Then, $N^{(q)} = 2K^{(q)}$ is the total number of clusters in such a parallel-randomized CRT study. Each cluster contains $m^{(q)}$ (simply $m$ if we ignore multiple comparisons) individuals indexed as $j = 1, \ldots, m^{(q)}$ in each cluster. We assume that all clusters are of the same size, that is, $m_q$ is equivalent in all clusters. The outcome vector for outcome $q$ is $Y_{q,ikj}$. In the case of two co-primary outcomes, the outcome vectors are $Y_{1,ikj}$ and $Y_{2,ikj}$, which we will refer to as $Y_1$ and $Y_2$, when it is possible to do so.

Consider $(Z_{1-\alpha/2} + Z_\beta)^2$ where $Z_{1-\alpha/2}$ is the $\left(1 - \frac{\alpha}{2}\right) \times 100^{th}$ lower percentile of the standard normal distribution with error rate $\alpha$, and $Z_\beta$ is the critical value corresponding to $\beta$. So, ignoring multiple testing, for example, at 80% power and a significance level of 5%, we have $(Z_{1-\alpha/2} + Z_\beta)^2 = (1.96 + 0.84)^2 = 7.84$. Then, in a CRT setting for difference in means for a single primary outcome $q$, the cluster size $m^{(q)}$ and the number of clusters in each arm $K^{(q)}$ has been given by the following equations[25]:

$$m^{(q)} = \frac{2(Z_{1-\alpha/2} + Z_\beta)^2 \sigma_q^2 \left(1 - \rho_0^{(q)}\right)}{\left(\beta_q^*\right)^2 K^{(q)} - 2(Z_{1-\alpha/2} + Z_\beta)^2 \sigma_q^2 \rho_0^{(q)}}, \tag{1}$$

$$K^{(q)} = \frac{2(Z_{1-\alpha/2} + Z_\beta)^2 \sigma_q^2 \left[1 + (m^{(q)} - 1)\rho_0^{(q)}\right]}{m^{(q)}(\beta_q^*)^2}, \quad (2)$$

where, for a given outcome $q = 1, \ldots, Q$, $\sigma_q^2$ is the total variance for outcome $q$, $\beta_q^*$ is the estimated treatment effect, and $\rho_0^{(q)}$ is outcome $q$'s intraclass correlation coefficient (ICC). The noncentrality parameter has been given by[25]:

$$\lambda^{(q)} = \frac{(\beta_q^*)^2}{2\frac{\sigma_q^2}{K^{(q)}m^{(q)}}\left[1 + (m^{(q)} - 1)\rho_0^{(q)}\right]}. \quad (3)$$

The power based on outcome $q$, denoted as $\pi_q$, for a two-sided test is calculated as $\pi = 1 - Pr(Z^2 \leq \chi^2(1)|H_A) = 1 - Pr(\chi^2(1,\lambda) \leq 3.84)$ for $\alpha = 5\%$. Here, $\chi^2(1)$ refers to the central $\chi^2$-distribution with 1-DF, and $\chi^2(1,\lambda)$ refers to the noncentral $\chi^2$-distribution with 1-DF. These expressions can be used also for binary outcomes; in the binary case, we simply use the variance for the difference between two binomial proportions from independent data instead of the variance for the difference between two means (namely $2\sigma_1^2$).

Although in our experience we have not seen this, there is sometimes a need to utilize unequal treatment allocation. As noted by Donner et al., Equation **(2)** can be adopted to incorporate a non-equal treatment allocation ratio, $r$.[25] Here, there are $N^{(q)} = K_1^{(q)} + K_2^{(q)}$ clusters total in the study. Then, we have the following expressions for the number of clusters in the treatment group, $K_1^{(q)}$, and in the control group, $K_2^{(q)}$:

$$K_1^{(q)} = \frac{\left(1 + \frac{1}{r}\right)(Z_{1-\alpha/2} + Z_\beta)^2 \sigma_q^2 \left[1 + (m^{(q)} - 1)\rho_0^{(q)}\right]}{m^{(q)}(\beta_q^*)^2},$$

$$K_2^{(q)} = \frac{(1 + r)(Z_{1-\alpha/2} + Z_\beta)^2 \sigma_q^2 \left[1 + (m^{(q)} - 1)\rho_0^{(q)}\right]}{m^{(q)}(\beta_q^*)^2}. \quad (4)$$

Note that $K_1^{(q)} = \frac{1}{2}K^{(q)}\left(1 + \frac{1}{r}\right)$ and $K_2^{(q)} = \frac{1}{2}K^{(q)}(1 + r)$.[25] Then the noncentrality parameter and cluster size under unequal treatment allocation for a given outcome $q$ is as follows:

$$\lambda^{(q)} = \frac{m^{(q)}(\beta_q^*)^2 K_1^{(q)}}{\left(1 + \frac{1}{r}\right)\sigma_q^2 \left[1 + (m^{(q)} - 1)\rho_0^{(q)}\right]}, \quad (5)$$

$$m^{(q)} = \frac{\left(1 + \frac{1}{r}\right)(Z_{1-\alpha/2} + Z_\beta)^2 \sigma_q^2 \left(1 - \rho_0^{(q)}\right)}{K_1^{(q)}(\beta_q^*)^2 - \left(1 + \frac{1}{r}\right)(Z_{1-\alpha/2} + Z_\beta)^2 \sigma_q^2 \rho_0^{(q)}}. \quad (6)$$

Next, we describe a hybrid CRT study that will illustrate how the possible study design methods for hybrid type 2 trials can be used.

### 1.8 Illustrative Example: Kaiser Implementation Strategy Bundle for Blood Pressure Control - CIRCL Hybrid Study

To illustrate the procedures and implications of the methodologies presented in the following sections, we utilize design input parameters motivated by a study being conducted by the Community Intervention to Reduce CardiovascuLar Disease in Chicago (CIRCL-Chicago) Implementation Research Center.[16,17] CIRCL-Chicago is a study to assess the impact of adapting a community-driven Kaiser implementation strategy bundle in Chicago to improve hypertension control.[16,17] The goal is to evaluate the use of practice facilitation (PF), which refers to supportive and collaborative approaches used to help healthcare practices, clinics, and providers improve patient care quality. The use of PF in this study is to help community health centers better utilize the community-adapted Kaiser bundle—a multilevel intervention, the adapted version of which is further described in Smith et al.[17]. The primary aim is to compare the effectiveness and implementation of the Kaiser bundle with PF (experimental group), compared to the implementation of a Kaiser bundle without PF (control group). There are $K = 15$ clinics in each treatment arm, with $m = 300$ patients in each clinic. The authors label this study is a hybrid type 3 study due to the nature of the research questions: the implementation outcome (reach) is the primary outcome given the strong evidence of effectiveness of the Kaiser bundle. Thus, all formal power calculations for this real-world study were based upon this endpoint. For the purposes of this article, we treat CIRCL-Chicago as a hybrid type 2 study to illustrate how input parameters available at the design stage could inform a similar hybrid type 2 study with characteristics similar to CIRCL-Chicago, in which the effectiveness and implementation outcomes were considered co-primary endpoints.

The first primary outcome, $Y_1$, is the proportion of patients with controlled blood pressure (BP) (yes/no). The second primary outcome, $Y_2$, is reach, which is defined as the proportion of patients, among those who were eligible, who received the Kaiser bundle (yes/no). These two metrics will be compared between the experimental and control group. The parameters assumed by the study investigators at the design stage will be used to calculate the study design specifications are shown in Table 2. Note that the total variances of $Y_1$ and $Y_2$ were adjusted to account for clustering, and these calculations based upon the data availability are shown in Appendix B.

## 2. Methods for Hybrid Type 2 Studies using Cluster Randomized Trial Designs

### 2.1 P-Value Adjustments for Multiple Testing

A popular group of methods that can be considered for the design and analysis of hybrid type 2 studies are p-value adjustments. To prevent inflation of the overall Type I error rate in a setting with multiple primary outcomes, adjustments to p-values are often utilized. Several p-value adjustment strategies can be applied to the case of two primary outcomes, the Bonferroni correction,[26] the Sidak method,[27] and the D/AP approach.[28] The p-value adjustment methods test the disjunctive hypothesis:

$$H_0: \beta_1^* = 0 \text{ and } \beta_2^* = 0 \quad \text{vs.} \quad H_A: \beta_1^* \neq 0 \text{ or } \beta_2^* \neq 0.$$

Here, we reject the null hypothesis if the intervention is found to have a significant effect on one or both of the outcomes, and we fail to reject the null only if the intervention does not have a significant effect on both outcomes. This hypothesis setup is synonymous to the hypothesis setup described in Vickerstaff et al., and is written as $H(Q) = \cap_{q=1}^{Q} H_q$, where $H_q$ is the null hypothesis for the $q$th outcome.[29] This overall null hypothesis is the case where there is no intervention effect on any of the outcomes, and so it can be seen that p-value adjustment methods make use of disjunctive power.

### 2.1.1. Bonferroni Correction

For two primary outcomes, the Bonferroni correction method preserves an overall family-wise Type I error rate of 5% by assigning a significance level of 0.025 (0.05/2) to each individual test. Recall that the family-wise Type I error rate is the probability of making one or more Type I errors when performing multiple testing. The overall Type I error rate does not necessarily need to be divided evenly between the two outcomes, and can be flexible in its division of the error rate depending on the setting. For example, if the effectiveness outcome of a hybrid type 2 study is more important to the given research question and is thought to be more difficult to power, then it may be assigned 0.03 while the implementation outcome could be assigned 0.02. However, it is important to note that these assignments must be made before the trial is conducted. The Bonferroni correction approach is a more conservative strategy than the Sidak and D/AP methods, but maintains similar levels of power when the strength of the correlation increases.[29]

### 2.1.2. Sidak Method

The Sidak method adjusts the standard p-value to obtain an overall 5% rejection rate,[27] using the laws of probability. In this method, the p-value is defined as $p^{Sdk} = 1 - (1-p)^Q$ where again $Q = 2$ is the total number of primary outcomes. As shown by Vickerstaff et al., with an unadjusted significance level $\alpha$, the Sidak adjusted significance level can expressed as $\alpha_{Sdk} = 1 - (1-\alpha)^{\frac{1}{Q}}$, which reduces to

$$\alpha_{Sdk} = 1 - \sqrt{(1-\alpha)}, \qquad (7)$$

when $Q = 2$, as shown here (see Appendix B.4 for details). These methods assume that the outcomes are independent. When the desired marginal p-value for a test is, for example, $\alpha = 0.05$, then $p^{Sdk} = 1 - (1 - 0.05)^2 = 0.0975$ to reject the null for each of the two tests separately. Equation **(7)** shows the Sidak adjusted Type I error rate. For the full derivation, please see Appendix B. When $Q = 2$ and $\alpha = 0.05$, as is usually the case, we obtain $\alpha_{Sdk} = 1 - (1 - 0.05)^{1/2} = 0.0253$. Note that this method does not take into account the correlation between the outcomes, but the Dubey/Armitage-Parmer (D/AP) method is an extension of the Sidak method that does take this correlation into account.[29]

### 2.1.3. D/AP Approach

The D/AP approach is an ad hoc method based that revises Sidak method to account for the correlation between multiple outcomes.[28] For hypothesis testing, the adjusted p-value is $p_q^{DAP} = 1 - (1-p)^{M_q}$ where $M_q = Q^{1-\rho(Y_q, Y_{-q})}$ is the multiple partial correlation coefficient between the $q$th outcome and the remaining outcomes, and $p$ is the marginal p-value, and $Q = 2$ in a hybrid type 2 study. Thus, $\rho(Y_q, Y_{-q})$ is the previously defined $\rho_2^{(1,2)}$, the correlation between the two outcomes for the same individual Type I error rate. At the design phase, we wish to solve for the Type I error rate, which is expressed as $\alpha_{DAP,2} = 1 - (1-\alpha)^{1/M_q}$ where $M_q = Q^{1-\rho_2^{(1,2)}}$.

In Appendix B.5, we prove that $\alpha^{Bonferroni} < \alpha^{Sidak} < \alpha^{DAP}$ for two or more outcomes. So, the D/AP method is less conservative than the Sidak method, which is less conservative than the Bonferroni method. For example, if $\rho_2^{(1,2)} = 0.5$, then for $\alpha = 0.05$, $\alpha_{DAP} = 0.0356$, which is less conservative than $\alpha^{Sdk} = 0.0253$ and $\alpha_{Bonf} = 0.025$. It is important to note that if the two primary outcomes are uncorrelated (i.e. $\rho_2^{(1,2)} = 0$), the D/AP $\alpha$-value reduces to the Bonferroni $\alpha$-value. Figure 1 illustrates the comparison between the D/AP and Bonferroni significance levels across different correlations for $Y_1$ and $Y_2$ in a hybrid type 2 design setting. In general, for setups with $Q$ outcomes, the $q^{th}$ adjusted D/AP p-value uses the correlation between the $q^{th}$ outcome and the remaining $Q-1$ outcomes. To obtain this, the square root of $R^2$ from a linear regression with the $q^{th}$ variable as the outcome and the remaining $Q-1$ variables as the predictors, denoted $R^2(q)$, may be used, such that $M_q = Q^{1-\sqrt{R^2(q)}}$.[29]

These adjustment methods are explored in simulation studies in Vickerstaff et al. in settings where all outcomes are continuous.[29] However, many studies including hybrid type 2 studies may utilize two binary outcomes, or a mixture of a binary and continuous outcome. P-value adjustment methods have not been thoroughly investigated in these settings. The correlation between two binary variables, or a binary and continuous variable, may lead to less extreme differences between these adjustment methods as compared to the case of two continuous outcomes.[29] Thus, additional work is needed in order to understand the performance of these methods more generally.

### 2.1.4. Power and Sample Size Calculations with P-value Adjustment

Power calculations for type 2 hybrid studies with p-value adjustments require integration over the noncentral $\chi^2$-distribution, but with a lower limit of integration, i.e. an alternate critical value where $\pi = 1 - \Pr(\chi_1^2(\lambda) \leq \text{Critical Value})$. For a 5% error rate, the critical value would be 3.84 in the standard setting with a single outcome. However, here, we revise the limit of integration to correspond to the new adjusted error rate in order to allow for two comparisons and a 5% family-wise error rate. Table 3 compares the critical values and noncentrality parameters for each p-value adjustment method outlined in this section, where the $\alpha_{DAP}$-level is calculated using different values of $\rho_1^{(1,2)}$.

To calculate the power of a hybrid type 2 CRT with a p-value adjustment, we have noncentrality parameter $\lambda$, which is a function of design features, including the number of clusters in each treatment arm ($K$), the number of individuals in each cluster ($m$), the overall variance of the outcome variables ($\sigma_1^2$ and $\sigma_2^2$), and the correlation coefficients ($\rho_0^{(1)}$ and $\rho_0^{(2)}$). Then, $\pi^{(q)} = 1 - \int_{CV}^{\infty} \chi^2(x; 1, \lambda^{(q)}) dx$, $q = 1, 2$ where $\lambda^{(q)}$ is given by:

$$\lambda^{(1)} = \frac{(\beta_1^*)^2}{2 \frac{\sigma_1^2}{Km}[1 + (m-1)\rho_0^{(1)}]}, \quad \lambda^{(2)} = \frac{(\beta_2^*)^2}{2 \frac{\sigma_2^2}{Km}[1 + (m-1)\rho_0^{(2)}]}.$$

The values of $\lambda^{(q)}$ are used to calculate the power of the study with the desired critical value depending on the p-value adjustment we wish to implement, giving us the statistical power based on outcomes 1 and 2, namely $\pi^{(1)}$ and $\pi^{(2)}$. The final power is the minimum of these two values, $\pi = \min(\pi^{(1)}, \pi^{(2)})$.

The required number of subjects within each cluster, $m$, to obtain a power of 80% is given by the following equations:

$$m^{(1)} = \frac{2(Z_{1-\alpha/2} + Z_\beta)^2 \sigma_1^2 (1 - \rho_0^{(1)})}{(\beta_1^*)^2 K - 2(Z_{1-\alpha/2} + Z_\beta)^2 \sigma_1^2 \rho_0^{(1)}}, \quad m^{(2)} = \frac{2(Z_{1-\alpha/2} + Z_\beta)^2 \sigma_2^2 (1 - \rho_0^{(2)})}{(\beta_2^*)^2 K - 2(Z_{1-\alpha/2} + Z_\beta)^2 \sigma_2^2 \rho_0^{(2)}},$$

where $m^{(1)}$ is the number subjects per cluster needed to obtain power for outcome 1, and $m^{(2)}$ applies to outcome 2. To guarantee 80% power for both outcomes, the final value is $m = \max(m^{(1)}, m^{(2)})$. To calculate the required number of clusters, $K$, needed in our study for 80% power, we use the following formulas:

$$K^{(1)} = \frac{2(Z_{1-\alpha/2} + Z_\beta)^2 \sigma_1^2 [1 + (m-1)\rho_0^{(1)}]}{m(\beta_1^*)^2}, K^{(2)} = \frac{2(Z_{1-\alpha/2} + Z_\beta)^2 \sigma_2^2 [1 + (m-1)\rho_0^{(2)}]}{m(\beta_2^*)^2},$$

where $K^{(1)}$ corresponds to outcome 1, and $K^{(2)}$ corresponds to outcome 2. To guarantee the desired power, $K = \max(K^{(1)}, K^{(2)})$. In calculations of both $K$ and $m$, the result will differ based on what p-value adjustment method we use to calculate $\lambda^{(q)}$. To account for unequal allocation, Equations **(4)**, **(5)**, and **(6)** can be used directly for this method.

## 2.2 Combined Outcomes Method

### 2.2.1. Types of combined outcomes

Another way of handling two outcomes is to combine them into a single outcome. There are three main types of composite endpoints that have been proposed as primary endpoints.[30] The first is a total score based on a specified rating scale. This may be a sum or average of scores from the same scale. This approach is suitable when the scale on which the total score is based has variability and reliability that is well established.[30] It is also helpful when the components of this composite score are not highly correlated; otherwise, the use of a single primary endpoint would contain virtually the same information. For example, common clinical endpoints in rheumatoid arthritis research are the American College of Rheumatology Criteria, ACR 20/50/70, which represents 20%, 50%, and 70% improvement on the number of swollen or tender joints.[31] Studies such as a clinical trial by Lipsky et al. examining the impact of Infliximab and Methotrexate on ACR 20/50/70 are thus evaluating a composite endpoint comprised of all of the joints that are symptomatic.[32] Naturally, these endpoints are dependent and related, so using a total score composite endpoint in settings such as this is fitting.

Another type of composite endpoint is a composite summary event count, which is computed by summing the number of predefined set of events that occur after a specified period of treatment or follow-up. Alternatively, it can be coded as 1 if any of the events occur, and 0 otherwise. For this endpoint, the events come from a predefined set of events relevant to the treatment and disease.[30] Chi et al. presents an example of a study that uses this type of endpoint where the composite endpoint was a six month failure rate, where a failure was defined as any of the three events taking place: biopsy-confirmed acute rejection, graft loss, or death.[30,33] The third type of composite endpoint is similar (and has been thought to be indistinct from the second type of composite endpoint), which is when multiple binary events or failure times are combined to form one composite event endpoint. This type of endpoint has been utilized in time event analysis in cardiovascular and chronic disease research; for example, a mortality outcome may be combined with a reduction in risk of acute myocardial infarction, stroke, or the need for a medical procedure.[30] The use of composite events relies on each event being clinically relevant.

Overall, combined outcomes can be an appropriate method for similar outcomes. In particular, composite endpoints help avoid multiplicity when there are several endpoints that similarly and sufficiently express the treatment effect. The use of composite endpoints allows for the inclusion of multiple relevant endpoints without increasing the sample size to allow for multiple testing.[30] A sum or average of combined outcome is most interpretable when all outcomes are on a similar scale. The outcomes must be similar in nature, clinically relevant, and meaningful. Chi et al. notes that these requirements are important because a significant result from the entire composite outcome could reflect a significant result from only one of the events or any combination of them, and so the significant result should be sufficient to support evidence of intervention efficacy if only one component is contributing overall. Thus, it is important to strike a balance by including outcomes within the composite outcome that add meaningful information, yet are also related to the same topic.

## 2.2.2. Power and Sample Size Calculations with Combined Outcomes

In a hybrid type 2 scenario, we have two primary outcomes, $Y_1$ and $Y_2$. Then, the combined outcome can be defined as the summation of the two outcomes: $Y_c = Y_1 + Y_2$. Here, $Y_c$ encapsulates both primary outcomes, where an increase in either of the outcomes will contribute to a larger value of $Y_c$. The parameter of interest for the combined outcome, $Y_c$, is $\beta_c^*$, and the hypothesis test for $\beta_c^*$ is:

$$H_0: \beta_c^* = 0 \quad \text{vs.} \quad H_A: \beta_c^* \neq 0.$$

Equations **(1)**, **(2)**, and **(3)** can be modified to apply to the combined outcome parameters. Thus, the noncentrality parameter for power calculations, the number of clusters in each treatment group (K), and the cluster size (m), can be calculated as:

$$\lambda = \frac{(\beta_c^*)^2}{2\frac{\sigma_c^2}{Km}\left[1 + (m-1)\rho_0^{(c)}\right]}, \quad K = \frac{2(Z_{1-\alpha/2} + Z_\beta)^2 \sigma_c^2 \left[1 + (m-1)\rho_0^{(c)}\right]}{m(\beta_c^*)^2},$$

$$m = \frac{2(Z_{1-\alpha/2} + Z_\beta)^2 \sigma_c^2 \left(1 - \rho_0^{(c)}\right)}{(\beta_c^*)^2 K - 2(Z_{1-\alpha/2} + Z_\beta)^2 \sigma_c^2 \rho_0^{(c)}},$$

where $\sigma_c^2$ is the total variance of the combined outcome, and $\rho_0^{(c)}$ is the endpoint specific ICC for the combined outcome. If preliminary or external data are available from which $Y_c$ can be calculated, $\rho_0^{(c)}$ can be directly calculated. Otherwise, $\rho_0^{(c)}$ can be approximated using $\rho_0^{(1)}$ and $\rho_0^{(2)}$, as

$$\rho_0^{(c)} = \frac{\rho_0^{(1)} \sigma_1^2 + \rho_0^{(2)} \sigma_2^2 + 2\rho_1^{(1,2)} \sigma_1 \sigma_2}{\sigma_1^2 + \sigma_2^2 + 2\rho_2^{(1,2)} \sigma_1 \sigma_2}. \quad (8)$$

Similarly, the total variance of the combined outcome can be estimated directly from pilot or preliminary data, or can be calculated from information about the variances of outcomes 1 and 2 as follows:

$$\sigma_c^2 = \sigma_1^2 + \sigma_2^2 + 2\rho_2^{(1,2)} \sigma_1 \sigma_2. \quad (9)$$

For the derivations of the total variance and endpoint specific ICC, see Appendix B. This method can be extended to unequal treatment allocation by using Equations **(4)**, **(5)**, and **(6)**, while using the combined outcome treatment effect ($\beta_c^*$), variance ($\sigma_c^2$), and ICC ($\rho_0^{(c)}$).

## 2.3 Single 1-DF Combined Test for Two Outcomes

In a one degree of freedom test for a hybrid type 2 scenario in this approach, the two separate test statistics are weighted to create a single test statistic. Here, we must extend this test originally proposed by Pocock et al. (1987) and O'Brien et al. (1984) to the CRT setting.[34,35] The general setup for $Q$ endpoints not restricted to $Q = 2$, as shown by Pocock et. al. (1987), but we focus on the case where $Q = 2$. For two primary outcomes, we define the test statistic as[34,35]

$$Z^2_{combined} = \left[\frac{Z_1 + Z_2}{\sqrt{2(1 + \text{Corr}(Z_1, Z_2))}}\right]^2 = \frac{(Z_1 + Z_2)^2}{2\left(1 + \frac{\left(\rho_2^{(1,2)} + (m-1)\rho_1^{(1,2)}\right)}{\sqrt{\left(1 + (m-1)\rho_0^{(1)}\right)\left(1 + (m-1)\rho_0^{(2)}\right)}}\right)},$$

where $Corr(Z_1, Z_2)$ is the correlation of the two test statistics, $Z_1$ and $Z_2$, defined below, and all other parameters defined as in Table 2. Under the null, the two test statistics are each distributed $\mathcal{N}(0,1)$. Now, define $\beta^* = (\beta_1^*, \beta_2^*)$ as the mean difference of the primary outcomes, $Y_1$ and $Y_2$, between treatment group 1 and treatment group 2. Then, under equal treatment allocation, the numerator of the test statistic is the sum of the two test statistics,

$$Z_1^2 = \frac{(\beta_1^*)^2}{\frac{2\sigma_1^2}{Km}[1 + (m-1)\rho_0^{(1)}]}, \quad Z_2^2 = \frac{(\beta_2^*)^2}{\frac{2\sigma_2^2}{Km}[1 + (m-1)\rho_0^{(2)}]}.$$

The test statistic, $Z^2_{combined}$, for this scenario when there are $Q = 2$ endpoints is the following:

$$Z^2_{combined} = \frac{\left(\sqrt{\frac{(\beta_1^*)^2}{\frac{2\sigma_1^2}{Km}[1 + (m-1)\rho_0^{(1)}]}} + \sqrt{\frac{(\beta_2^*)^2}{\frac{2\sigma_2^2}{Km}[1 + (m-1)\rho_0^{(2)}]}}\right)^2}{2\left(1 + \frac{\left(\rho_2^{(1,2)} + (m-1)\rho_1^{(1,2)}\right)}{\sqrt{\left(1 + (m-1)\rho_0^{(1)}\right)\left(1 + (m-1)\rho_0^{(2)}\right)}}\right)} \sim \chi^2(1),$$

with 1-DF under $H_0$. This method can be generalized to account for unequal treatment allocation; the full derivation for the generalized case is shown in Appendix A.

The p-value for this single 1-DF combined test can be calculated by $1 - \Pr(\chi^2(1) \leq Z^2_{combined})$. In this setting, the hypothesis we are testing is

$$H_0: \beta_1^* + \beta_2^* = 0 \text{ vs. } H_A: \beta_1^* + \beta_2^* \neq 0,$$

and with a significant p-value, we would conclude that the intervention had an effect on at least one of the outcomes. To eliminate the possibility that $\beta_1^* = -\beta_2^*$, it is assumed that both elements of $\beta^*$ are in the same direction, and data can be transformed to ensure that this is the case. Then, under this requirement that $\beta_1^*$ has the same sign as $\beta_2^*$, which is a reasonable for two co-primary endpoints of the same intervention, $\beta_1^* + \beta_2^* = 0$ only if $\beta_1^* = 0$ and $\beta_2^* = 0$. The alternative is the case in which one or both of the effects is not equal to zero, also written as $\beta_1^* \neq 0$ or $\beta_2^* \neq 0$, and this test is therefore a disjunctive test.

Continuing with the study design of a hybrid type 2 study using a CRT design, the noncentrality parameter, $\lambda$, for this test is approximated by the combined test statistic:

$$\lambda = \left[ \frac{\sqrt{\frac{(\beta_1^*)^2}{\frac{2\sigma_1^2}{Km}[1+(m-1)\rho_0^{(1)}]}} + \sqrt{\frac{(\beta_2^*)^2}{\frac{2\sigma_2^2}{Km}[1+(m-1)\rho_0^{(2)}]}}}{\sqrt{2\left(1 + \frac{\left(\rho_2^{(1,2)} + (m-1)\rho_1^{(1,2)}\right)}{\sqrt{\left(1+(m-1)\rho_0^{(1)}\right)\left(1+(m-1)\rho_0^{(2)}\right)}}\right)}} \right]^2. \qquad (10)$$

Note that the variance for both outcomes, $\sigma_1^2$ and $\sigma_2^2$, are included in this equation for the noncentrality parameter used for study design, as well as the correlation coefficients $\rho_0^{(1)}$, $\rho_0^{(2)}$, $\rho_1^{(1,2)}$, and $\rho_2^{(1,2)}$. So, power calculations based on this test require these parameters as well as $K$, $m$, and the intervention effects of the two outcomes. One can use the noncentrality parameter to calculate the statistical power.

Solving Equation **(10)** for $K$ will allow us to find the required number of clusters in each treatment group, which is shown as Equation **(11)**. Solving for $m$ will allow us to find the required cluster size, and can be calculated with an equation solver in R. The equation for $K$ is

$$K = \frac{2(Z_{1-\alpha/2} + Z_\beta)^2 \left(1 + \frac{\left(\rho_2^{(1,2)} + (m-1)\rho_1^{(1,2)}\right)}{\sqrt{\left(1+(m-1)\rho_0^{(1)}\right)\left(1+(m-1)\rho_0^{(2)}\right)}}\right)}{\left[\sqrt{\frac{(\beta_1^*)^2}{\frac{2\sigma_1^2}{m}[1+(m-1)\rho_0^{(1)}]}} + \sqrt{\frac{(\beta_2^*)^2}{\frac{2\sigma_2^2}{m}[1+(m-1)\rho_0^{(2)}]}}\right]^2}. \qquad (11)$$

### 2.4 Disjunctive 2-DF Test for Two Outcomes

In this test for hybrid type 2 studies, we simultaneously test both outcomes for any departure from the null hypothesis of no intervention effect across any outcome. The general linear hypothesis of interest can be written as follows:

$$H_0: \boldsymbol{L\beta^*} = \boldsymbol{0} \text{ vs. } H_A: \boldsymbol{L\beta^*} \neq \boldsymbol{0},$$

where $\boldsymbol{L}$ is a contrast matrix whose rows represent linearly independent hypotheses concerning the treatment effect parameter vector, $\boldsymbol{\beta^*}$. Here, we have two hypotheses and $Q = 2$ outcomes, so $\boldsymbol{L}$ is a $2 \times 2$ matrix and is $\boldsymbol{L} = \begin{bmatrix} 1 & 0 \\ 0 & 1 \end{bmatrix}$. Then, the hypothesis is written as

$$H_0: \beta_1^* = 0 \text{ and } \beta_2^* = 0 \quad \text{vs.} \quad H_A: \beta_1^* \neq 0 \text{ or } \beta_2^* \neq 0.$$

Then, an overall test statistic for $H_0$ in a CRT setting has the following form under the null[36]:

$$F^* = K(\boldsymbol{L\beta^*})^T \left(\boldsymbol{L} \Omega_{\beta^*} \boldsymbol{L}^T\right)^{-1} \boldsymbol{L\beta^*},$$

where $\Omega_{\beta^*}$ is a $2 \times 2$ variance-covariance matrix for the treatment effect estimate $\beta^*$. Then, for the hybrid type 2 design with equal treatment allocation (where the variance of the treatment allocation is $\sigma_z^2 = 1/4$), we have

$$\omega_1^2 = \frac{4\sigma_1^2[1+(m-1)\rho_0^{(1)}]}{m}, \quad \omega_2^2 = \frac{4\sigma_2^2[1+(m-1)\rho_0^{(2)}]}{m},$$

$$\omega_{1,2} = \frac{4\sigma_1\sigma_2\,[\rho_2^{(1,2)} + (m-1)\rho_1^{(1,2)}]}{m},$$

where $\omega_1^2$ and $\omega_2^2$ are the elements on the main diagonal of the $2 \times 2$ variance-covariance matrix, and $\omega_{qq'}$ is the element of the off-diagonal.[36] Each term above contains $\sigma_z^2$ in the denominator, which results in the numerator containing a 4 under equal allocation. As always, in the design of hybrid type 2 studies using CRTs, these parameters would need to be ascertained from pilot studies, or the literature, but in an analysis setting, these would be estimated from the data. The noncentrality parameter is defined as $\lambda = 2K(L\beta^*)^T\left(L\,\Omega_{\beta^*}L^T\right)^{-1}L\beta^*$. For two primary outcomes and equal intervention allocation, the expression simplifies (Appendix B.6), and the noncentrality parameter is

$$\lambda = \left[\frac{Km[(\beta_1^*)^2\sigma_2^2\,VIF_2 - 2\beta_1^*\beta_2^*\sigma_1\sigma_2 VIF_{12} + (\beta_2^*)^2\sigma_1^2 VIF_1]}{2\sigma_1^2\sigma_2^2[VIF_1\,VIF_2 - VIF_{12}^2]}\right], \quad (12)$$

where $VIF_1 = 1 + (m-1)\rho_0^{(1)}$, $VIF_2 = 1 + (m-1)\rho_0^{(2)}$, and $VIF_{12} = \rho_2^{(1,2)} + (m-1)\rho_1^{(1,2)}$. We note that in a hybrid type 2 study using this test, four correlations must be specified, defined in Table 2, namely $\rho_0^{(1)}$, $\rho_0^{(2)}$, $\rho_1^{(1,2)}$, and $\rho_2^{(1,2)}$. Then, for a prespecified Type I error rate $\alpha$, the power is

$$\pi = \int_{F_{1-\alpha}(S,2K-S-Q)}^{\infty} f(x;\lambda,2,2K-2-Q)dx,$$

where $F_{1-\alpha}(2, 2K-2-Q)$ is the critical value of the central $F(2, 2K-2-Q)$ distribution, and $f(x;\lambda,2,2K-2-Q)$ is the probability density function of the noncentral $F(\lambda, 2, 2K-2-Q)$ distribution.[36] Since the degrees of freedom is a function of the number of clusters, to solve for the number of clusters in the treatment group, $K$, iterative numerical methods must be used.

A $\chi^2$-distribution can be used instead of the F-distribution when $K$ is large. As the number of clusters increases, the results from using the F-distribution and the $\chi^2$-distribution converge. More precisely, when $K \to \infty$, a F-distribution with noncentrality parameter $\lambda$ and numerator and denominator degrees of freedom $(2, 2K-2-Q)$ will converge to a $\chi^2$-distribution with noncentrality parameter $\lambda$ and 2 degrees of freedom.[37] This suggests that in the hybrid type 2 case when $Q = 2$, $F_{1-\alpha}(2, 2K-4)$ and $f(x;\lambda,2,2K-4)$ converges to $\chi_{1-\alpha}^2(2)$ and $\chi^2(x;2,\lambda)$, respectively, where $\chi_{1-\alpha}^2(2) = 5.99$ is the $(1-\alpha) \times 100$th lower percentile of a central $\chi^2$-distribution with 2 degrees of freedom, and $\chi^2(x;2,\lambda)$ is the probability density function of the noncentral $\chi^2$-distribution with 2 degrees of freedom and noncentrality parameter $\lambda$ defined above. It follows that the expression for power becomes

$$\pi = \int_{\chi_{1-\alpha}^2(2)}^{\infty} \chi^2(x;2,\lambda)dx,$$

where $\chi_{1-\alpha}^2(2) = 5.99$ if we set the significant level $\alpha$ to 5%. Since the degrees of freedom are not a function of $K$ for the $\chi^2$-distribution, we can solve Equation **(12)** for $K$ to obtain an expression. This gives the following equation:

$$K = \left[\frac{2(Z_{1-\alpha/2} + Z_\beta)^2\,\sigma_1^2\,\sigma_2^2\,[VIF_1 VIF_2 - VIF_{12}^2]}{m[(\beta_1^*)^2\,\sigma_2^2\,VIF_2 - 2\,\beta_1^*\,\beta_2^*\,\sigma_1\,\sigma_2\,VIF_{12} + (\beta_2^*)^2\,\sigma_1^2\,VIF_1]}\right]. \quad (13)$$

Recall that $VIF_1$, $VIF_2$, and $VIF_{12}$ are functions of $m$. To calculate $m$ for both the F-distribution and $\chi^2$-distribution, we can use an equation solver in R, such as "multiroot()" from the "rootSolve" package, to solve the non-linear equation for $m$ for a specified power level and number of clusters, $K$, using Equation **(13)**. As noted previously, the expression for $K$ simplifies nicely under equal treatment allocation, but this test also accommodates unequal treatment allocation. Under unequal treatment assignment, $\sigma_z^2 \neq 1/4$, and this will be reflected in the covariance matrix $\Omega_{\beta^*}$, thus affecting the design equations. For more details, please see Yang et al.[36] R software for both the equal and unequal treatment allocation case is available on the GitHub repository located at https://github.com/siyunyang/coprimary_CRT.

## 2.5 Conjunctive Intersection-Union Test

The conjunctive test is powered for the alternative that the intervention has an effect on both outcomes. This test is the only test currently available that tests whether or not the intervention has an effect on both outcomes rather than either outcome, which we believe is more closely aligned with what investigators using hybrid type 2 studies are interested in. This test specifies the following hypotheses:

$$H_0: \beta_1^* = 0 \text{ or } \beta_2^* = 0 \quad \text{vs.} \quad H_A: \beta_1^* \neq 0 \text{ and } \beta_2^* \neq 0.$$

This test cannot be written as a linear hypothesis of the standard form $\boldsymbol{L}\,\beta^*$. For testing $H_0$, Yang et al. consider the vector of Wald test statistics; for two outcomes, this is $\boldsymbol{\zeta} = (\zeta_1, \zeta_2)^T$, where $\zeta_q = \sqrt{K_1 + K_2}\beta_q^*/\omega_q$ and $\omega_q$ is the estimated standard error of the treatment effect estimator and $K_1 + K_2$ are the total clusters in the study ($2K$ under equal treatment allocation).[36] Here, $\beta_1^*$ and $\beta_2^*$ are being divided by the square root of the variances, so the effect sizes become standardized effect sizes. The test statistic vector can be expressed as a function of the VIF quantities, and under equal treatment allocation is

$$[\zeta_1, \zeta_2]^T = \left[\frac{\beta_1^*\sqrt{2K}}{\sqrt{\frac{4\sigma_1^2 VIF_1}{m}}}, \frac{\beta_2^*\sqrt{2K}}{\sqrt{\frac{4\sigma_2^2 VIF_2}{m}}}\right]^T. \tag{14}$$

In the context of a hybrid type 2 studies where $Q = 2$, the power is

$$\pi = \Pr\left(R = \bigcap_{q=1}^{2}\{\zeta_q > c_q\} \,|\, H_A\right) = \int_{c_1}^{\infty}\int_{c_2}^{\infty} f_{\boldsymbol{W}}(w_1, w_2)\, dw_1 dw_2,$$

where $R$ denotes the pre-specified rejection region, $c_1$ and $c_2$ are the corresponding endpoint specific critical values for rejection for outcomes 1 and 2 respectively, and $f_{\boldsymbol{W}(w_1,w_2)}$ is typically chosen to be a multivariate normal (MVN) density function for the two endpoint-specific Wald test statistics jointly under the alternative with mean vector $\sqrt{2K}\,(\beta_1^*/\omega_1, \beta_2^*/\omega_2)^T$ and covariance matrix $\Phi$.[36] The $2 \times 2$ correlation matrix, $\Phi$, is given by Yang et. al (2022), where the main diagonal is 1 and the off-diagonal is $\phi_{12} = \frac{\omega_{12}}{\omega_1 \omega_2}$, with $\omega_1^2$, $\omega_2^2$, and $\omega_{1,2}$ are as defined previously. As is the case for the disjunctive test, the adjustment for unequal treatment allocation comes from $\sigma_z^2$, and affects the design equations through $\Phi$.

Though a multivariate normal density function can be utilized for $f_{\boldsymbol{W}(w_1,w_2)}$, the authors note that a multivariate t-distribution can also be used to better control for Type I error in a setting where there is a small amount of clusters.[36] We denote the multivariate t-distribution density as $f_{T(\omega_1,\omega_2)}$ with $2K - 2Q = 2K - 4$ degrees of freedom to correct for finite sample bias. Then, a simple approach for finding the critical values is given by Yang et al., where $c_1 = c_2 = t_\alpha(2K - 2Q)$, where $t_\alpha(2K - 2Q)$ is the $(1 - \alpha)$ quantile of the univariate t-distribution.[36] Using the t-distribution, the power is

$$\pi = \int_{c_1}^{\infty} \int_{c_2}^{\infty} f_T(w_1, w_2; \boldsymbol{\zeta}, \boldsymbol{\Phi}, 2K - 4) \, dw_1 dw_2.$$

As was the case with the disjunctive test using the F-distribution, when using the t-distribution, the degrees of freedom is a function of the number of clusters. So, to solve for the number of clusters in the treatment group, $K$, iterative numerical methods are also used here. To calculate $m$, we use R software to solve for $m$ using the power equations above. R software has been developed for power and sample size calculations using the function "calPower_ttestIU()" from the GitHub repository available at https://github.com/siyunyang/coprimary_CRT. This software also allows for unequal treatment allocation, where the user can specify the proportion of clusters in the experimental group.

## 3. Illustrative Example: The CIRCL-Chicago Study

In this section, we show how various power and sample size calculations are done using the five methods described previously. All design methods used the relevant parameter values displayed in Table 2 for the CIRCL-Chicago study as provided by the investigators (co-authors JDS and YT). For Methods 1-4, we show the results when using a $\chi^2$-distribution, but results for power using the F-distribution are also shown in Table 5. For Method 5, we show results for both the MVN-distribution and t-distribution. The R code used in all power and sample size calculations for all five methods in this applied example is available in the code appendix on our GitHub repository, https://github.com/melodyaowen/Hybrid2DesignCIRCL. More details regarding the calculations are shown in Appendix D.

### 3.1 P-Value Adjustment Methods for CIRCL-Chicago Study

#### 3.1.1 Power

To calculate power based on the three p-value adjustment methods for the CIRCL-Chicago study, we first calculate the noncentrality parameter for the first outcome as follows:

$$\lambda^{(1)} = \frac{(0.1)^2}{2\frac{0.23}{15(300)}[1 + (300-1)0.025]} = 11.54.$$

Using the same formula for the second outcome, we have $\lambda^{(2)} = 10.62$. The critical values used to find the statistical power are calculated in R as $\chi^2_{1-\alpha}(1)$, where we have 1-DF and the value of $\alpha$ in this expression differs based on the p-value adjustment method. These critical values are displayed in Table 3. Power is calculated based on the three adjustment methods for both the effectiveness outcome ($Y_1$) and the implementation outcome ($Y_2$) separately. For example, for the Bonferroni adjustment, we have $\pi^{(1)}_{Bonf} = 1 - \Pr(\chi^2(1, \lambda^{(1)}) \leq 5.02) = 87.62\%$ and $\pi^{(2)}_{Bonf} = 1 - \Pr(\chi^2(1, \lambda^{(2)}) \leq 5.02) = 84.55\%$. Taking the minimum among the two outcomes for each method, the final results are $\pi_{Bonf} = 84.55\%$, $\pi_{Sidak} = 84.67\%$, and $\pi_{D/AP} = 84.98\%$.

#### 3.1.2 Number of Clusters (K)

To calculate the required number of clusters ($K$) for each study arm for fixed power using the three p-value adjustment methods, we take the maximum value of the two outcomes for each method. The noncentrality parameters that correspond to each p-value method for 80% power and overall p-value of 0.05 is the solution to $\zeta = \left(Z_{\alpha_{Adj.\ Method}} + Z_\beta\right)^2$ for the equation $80\% = 1 - \Pr(\chi^2(1,\zeta) \leq \text{Critical Value})$, where the critical values based on the adjustment method, as aforementioned, are shown in Table 3. The solution using the critical value for the Bonferroni method yields 9.51, the solution using the critical for the Sidak method yields 9.47, and the solution using the critical value of the D/AP method yields 9.39. Then, from Equation **(2)**, we can calculate $K$ for the two outcomes, $Y_1$ and $Y_2$, based on each of the three methods. For example, for the Bonferroni method for the first outcome, we have

$$K_{Bonf}^{(1)} = \frac{2(9.51)(0.23)[1 + (300-1)0.025]}{300(0.1)^2} = 12.35 \approx 13.$$

Using the same formula for the second outcome, we obtain $K_{Bonf}^{(2)} = 14$. The final result for the number of clusters in each treatment group based on the Bonferroni method is $K_{Bonf} = \max\left(K_{Bonf}^{(1)}, K_{Bonf}^{(2)}\right) = 14$. Following the same process for the Sidak and D/AP method, the final results these methods are $K_{Sidak} = 14$ and $K_{D/AP} = 14$.

### 3.1.3 Cluster Size (m)

Similarly, to calculate the required cluster size, $m$, using the three p-value adjustment methods, we take the maximum value of the two outcomes for each method. For example, using the noncentrality parameters calculated above and following Equation **(1)**, the cluster size for each outcome for the Bonferroni method for the first outcome is

$$m_{Bonf}^{(1)} = \frac{2(9.51)(0.23)(1 - 0.025)}{(0.1)^2 15 - 2(9.51)(0.23)(0.025)} = 104.76 \approx 105.$$

Using the same formula for the second outcome, we obtain $m_{Bonf}^{(2)} = 149$. The final result for the cluster size based on the Bonferroni method is $m_{Bonf} = \max\left(m_{Bonf}^{(1)}, m_{Bonf}^{(2)}\right) = 149$. Following the same process for the Sidak and D/AP method, the final results for these methods are $m_{Sidak} = 147$ and $m_{D/AP} = 141$. Table 4 summarizes the power and sample size estimates derived separately for each outcome.

## 3.2 Combined Outcomes Methods for CIRCL-Chicago Study

Recall that for this approach, we have two primary outcomes, $Y_1$ and $Y_2$, and we define the combined outcome as $Y_c = Y_1 + Y_2$. Here, an increase in the proportion of patients with controlled BP ($Y_1$) or an increase in the proportion of patients who received the Kaiser bundle among those who were eligible ($Y_2$), both contribute to larger values of $Y_c$.

From the study design parameters provided, we estimate the combined outcome intervention effect, $\beta_c^*$, as $\beta_1^* + \beta_1^* = 0.1 + 0.1 = 0.2$. We also calculate the total variance of the combined outcome, $\sigma_c^2$, using Equation **(9)**, which gives $\sigma_c^2 = 0.50$. Lastly, we can estimate $\rho_0^{(c)}$ using Equation **(8)**, the endpoint specific ICC for the combined outcome, which gives $\rho_0^{(c)} = 0.03$. With these results, we can calculate the power, number of clusters needed for each treatment arm, and cluster size needed for a new study based on this approach.

### 3.2.1 Power

To power a new study assuming parameters as shown in Table 2 and the parameters calculated above for the combined outcome approach, we use $\beta_c^*$ and its variance in the power equation for CRTs. The power of this study is calculated by first calculating the noncentrality parameter,

$$\lambda = \frac{(0.2)^2}{2\frac{0.50}{15(300)}[1 + (300-1)(0.03)]} = 16.42,$$

and then the power, $\pi$, is $\pi = 1 - Pr(\chi^2(1,\lambda) \leq 3.84) = 98.18\%$.

### 3.2.2 Number of Clusters (K)

To calculate the required number of clusters for pre-specified power, $\pi$, and numer of participants within each cluster, $m$, we use Equation **(2)** for calculating $K$ for a CRT under the parameters calculated previously, and $(Z_{1-\alpha/2} + Z_\beta)^2 = 7.85$ corresponding to $\pi = 80\%$ and $\alpha = 0.05$ with 1-DF. Using the equation it follows that $K = 7.17$, which we round up to $K = 8$.

### 3.2.3 Cluster Size (m)

To calculate the required cluster size for pre-specified power, $\pi$, and number of clusters in each treatment arm, $K$, we use Equation **(1)** for the combined outcomes approach under parameters calculated above from the input parameters, and $(Z_{1-\alpha/2} + Z_\beta)^2 = 7.85$ corresponding to $\pi = 80\%$ and $\alpha = 0.05$ for a noncentral $\chi^2$ with 1-DF. Then, using the equation it follows that $m = 22.42$, which we round up to $m = 23$.

## 3.3 Single 1-DF Combined Test Method with CIRCL-Chicago Study

### 3.3.1 Power

To calculate the power of a study under this single 1-DF combined test, we first calculate the noncentrality parameter with the parameters given by Table 2. Using Equation **(10)**, the noncentrality parameter is $\lambda = 16.30$. Using this quantity, the final study power based on this 1-DF combined test with $\alpha = 0.05$ is $\pi = 98.11\%$; that is, $Pr(\chi^2(1,\lambda) > 3.84) = 0.9811$.

### 3.3.2 Number of Clusters (K)

To calculate the required number of clusters per study arm in this setting, we again assume the parameters given by Table 2. For 80% statistical power and $\alpha = 0.05$, we have $(Z_{1-\alpha/2} + Z_\beta)^2 = 7.85$. Then, following Equation **(11)**, we obtain $K = 7.22$, which we round up to $K = 8$ required clusters per study arm.

### 3.3.3 Cluster Size (m)

To calculate the required cluster size for this test at a pre-specified power, $\pi$, and Type I error rate, $\alpha$, we solve Equation **(11)** for $m$. Using an equation solver in R, the solution to this equation yields $m = 22.69 \approx 23$.

## 3.4 Disjunctive 2-DF Test with CIRCL-Chicago Study

### 3.4.1 Power

Next, we find the power of a hybrid type 2 study under this test. To obtain $\lambda$, we first calculate the variance inflation factors. For example, $VIF_1$ is calculated as

$$VIF_1 = 1 + (m-1)\rho_0^{(1)} = 1 + (300-1)(0.025) = 8.48,$$

and similarly, we also have $VIF_2 = 8.48$ and $VIF_{12} = 3.04$. Then, plugging in all the required values into Equation **(12)**, we have

$$\lambda = \left[\frac{15(300)[(0.1)^2(0.25)(8.48) - 2(0.1)(0.1)\sqrt{0.23}\sqrt{0.25}(3.04) + (0.1)^2(0.23)(8.48)]}{2(0.23)(0.25)[(8.48)(8.48) - (3.04)^2]}\right],$$

giving us $\lambda = 16.32$, and using the critical value calculated in R as $\chi^2_{1-\alpha}(2) = 5.99$, we obtain the power as $\pi = 1 - \int_0^{5.99} f(x; 2, \lambda = 16.32) dx = 96.01\%$.

### 3.4.2 Number of Clusters (K)

Next, we find the number of clusters needed under this test for pre-specified power, $\pi$, and cluster size, $m$. For $\pi = 80\%$ power, $\alpha = 0.05$, and 2 degrees of freedom, we use $(Z_{1-\alpha/2} + Z_\beta)^2 = 9.63$. Then, plugging in the relevant values into Equation **(13)**, we have $K = 8.86$, which we round to $K = 9$ clusters per treatment group.

### 3.4.3 Cluster Size (m)

For $\pi = 80\%$ power and $\alpha = 0.05$, we again have $(Z_{1-\alpha/2} + Z_\beta)^2 = 9.63$. Then, for $K = 15$ clusters in each treatment group, we solve Equation **(12)** for $m$. Because the variance inflation factor terms are also functions of $m$, it is not possible to obtain a closed form solution for $m$. Thus, we use an equation solver in R to solve for $m$, and obtain that the required cluster size is $m = 33.84 \approx 34$. When the F-distribution is used instead, the upper bound of the integration for power (i.e. $\chi^2_{1-\alpha}(2) = 5.99$) and the term $(Z_{1-\alpha/2} + Z_\beta)^2 = 9.63$ change to be 3.37 and 10.84 respectively, since they instead are calculated using the F-distribution instead of the $\chi^2$-distribution.

### 3.5 Conjunctive Test with CIRCL-Chicago Study

To calculate power and sample size, we use the functions called "calPower_ttestIU" and "calSampleSize_ttestIU" in R, developed by Siyun Yang.[36] Under this test, the critical values are set to $c_1 = c_2 = t_\alpha(2K - 4)$, where $t_\alpha(2K - 4)$ is the $(1 - \alpha)$ quantile of the univariate t-distribution.[36] Yang et al. explains that this specification of the critical values ensure that the Type I error rate is strictly below $\alpha$ within the composite null space, which says that the treatment effect is zero for at least one outcome.[36]

Given the design parameters as specified previously, the test statistic vector is calculated using Equation **(14)** as $[\zeta_1, \zeta_2]^T = [3.40, 3.26]^T$. The critical value used in power calculations based on the t-distribution is $c_1 = c_2 = t_\alpha(2 \times 15 - 4) = 1.71$. Then, in this setting where the Type I error rate of 5%, $K = 15$, and $m = 300$, a power of $\pi = 89.92\%$ is obtained. If instead we would like to calculate the number of clusters per study arm needed when power is set to 80% and $m = 300$, using the same function in R, we see that $K = 12$ clusters per study arm are needed for this design. Lastly, if we would like to calculate the cluster size for a study with 80% power and $K = 15$, we find that we need $m = 86$ participants per cluster. If we wish to use the multivariate normal distribution, the critical value would be $c_1 = c_2 = 1.64$. Using this distribution, the statistical power is $\pi = 91.43\%$, the number of clusters per treatment group is $K = 11$, and the cluster size is $m = 74$. These results are also displayed in Table 5.

### 3.6 Summary of Results for CIRCL-Chicago Study

To summarize the performance of the various study design methods CIRCL-Chicago, Table 5 displays the power and sample sizes that each test yields when using the inputs given in Table 2. When comparing these five study design approaches in the context of the CIRCL-Chicago hybrid study, we see that the combined outcome approach yielded the highest statistical power. This is maybe because in combining the outcomes, we are essentially looking at only one outcome. We can reject the null when the treatment is effective for one or both outcomes, and also when the treatment is not effective on either outcome – but the combination of the two is effective. Though the combined outcome approach yielded higher power and is a popular approach, it is always important for investigators to verify that the hypothesis framework matches their research question and overall study goals. Similarly, the single 1-DF test also yielded very high statistical power and low sample size requirements. A non-linear relationship was found between the combined outcomes approach and the single 1-DF test. Appendix C shows that when the outcome specific ICCs and the variances for both outcomes are the same, these two methods are mathematically equivalent.

Next, the p-value adjustment methods yielded the lowest power compared to the other methods, and had the highest sample size requirements for achieving 80% power. The differences in the results between the three p-value adjustment methods are negligible, and this is because there are only two outcomes. As the number of outcomes increases, the difference in the results between the three methods will also increase. For example, for $Q = 2$ outcomes, $\alpha_{Bonf} = 0.025$, $\alpha_{Sidak} = 0.0253$, and $\alpha_{D/AP} = 0.0262$. When increasing the number of outcomes to $Q = 100$ (and keeping $\rho_2^{(1,2)} = 0.05$), these values become $\alpha_{Bonf} = 0.0005$, $\alpha_{Sidak} = 0.000513$, and $\alpha_{D/AP} = 0.00646$ and differ more greatly than in the case of two outcomes. P-value adjustment methods are popular, and it is important to remember that they do not require the intervention to be effective on both outcomes to reject the null hypothesis.

The disjunctive 2-DF test yielded the third highest power, lower than the combined outcomes approach and the single 1-DF combined test. This test uses a linear hypothesis framework. The conjunctive test yielded the second lowest statistical power, performing better than the p-value adjustment methods. This could be due to the fact that this test is powered for an alternative hypothesis that the intervention has an effect on both of the outcomes. Thus, the rejection region for this hypothesis test is smaller than the hypothesis framework where the intervention is only required to have an effect on one outcome to reject the null. It is advantageous because it tests whether or not the intervention has an effect on both of the outcomes rather than either outcome, but it does require a larger sample size.

## 4. Discussion

In this paper, new and existing statistical study design approaches and frameworks for handling two outcome variables in type 2 hybrid studies were described and compared. Namely, this paper discussed the p-value adjustment methods, combined outcomes approach, single 1-DF test, disjunctive 2-DF test, and conjunctive test. The combined outcomes approach and the single 1-DF test were extended to provide valid results for CRTs with two co-primary outcomes. Study design input parameters motivated by the CIRCL-Chicago hybrid effectiveness-implementation study[16,17] were used to show how these methods may be used to calculate the power, number of clusters, and cluster size required for a study, and the results from these methods for this setting were compared.

The conjunctive test yielded lower power compared to most of the other design methods for CIRCL-Chicago, but may be advantageous because it tests for an alternative hypothesis that requires the intervention to have an effect on both outcomes. However, the combined outcomes approach and single 1-DF test yield the highest power, and may be fitting in scenarios where resources restrict how many clusters or individuals can be included in a study. In particular, the single 1-DF test makes use of all of the correlations in a CRT, and they serve as weights for the overall test statistic, making this test a potential great alternative to the p-value adjustment methods, which don't make use of the inter- and intra-cluster correlations. Still, the p-value adjustment methods are easy to enact and are a popular design technique, and should not be discounted. Overall, when choosing a statistical design for hybrid studies, it is important to carefully consider the hypothesis framework that is best suited for the study's research questions, while also ensuring the study design is feasible in the real-world. These tests provide many options for researchers in the design of their hybrid 2 studies.

Important limitations exist in study design methodology for hybrid type 2 studies, one of which is the availability of software. PASS is a widely used software that allows users to conduct power calculations for various scenarios, including CRTs, but does not support all of the various design methods that are relevant in a co-primary endpoint setting. PASS is also not freely available. Some R packages exist such as the popular "pwr" package, but it does not take into account clustered data nor multiple outcomes; to our knowledge, no package currently exists in R that both handles CRT data with multiple primary outcomes. To address this issue, we have developed an R package called "crt2power". It is available on the Comprehensive R Archive Network (CRAN), and includes code for all of the possible design methods discussed in this paper, permitting researchers to easily apply these methods accurately.[38] A paper that introduces this package is currently being developed. There is also a need to examine these methods in many different scenarios in order to better understand their performance, and discern whether certain methods are better than others in particular design scenarios. Thus, this paper will also include a numerical study that will make use of this new software, and allow for more comprehensive comparisons between these study design methods.

Other gaps pertaining to the methodology also exist – the disjunctive 2-DF test, conjunctive test, and p-value adjustment methods have been extended to accommodate unequal cluster sizes, but the combined outcomes test and the single 1-DF test do not. Furthermore, we focus on two binary outcomes in our applied example, but future work could examine the impacts of differing outcome types (i.e. one continuous and one binary outcome). Future work could also examine hybrid 2 studies and stepped wedge designs. Lastly, small sample size correction in CRTs is important to consider, but outside the scope of this manuscript. Future research can build on previous work that has been done to account for this issue, and adapt it to hybrid 2 studies; we refer the reader to Leyrat et al., Thompson et al., and Taljaard et al., for more information.[39-41]

The rising popularity of hybrid studies could be attributed to their efficiency in bridging the gap between evaluating treatment and implementation strategy effectiveness. To harness the full potential of these studies accurately, particularly a type 2 study, it is crucial to bear in mind the statistical complexities we have explored, particularly those that arise with the presence of multiple primary outcomes and clustered data. Our aim with introducing and examining these study design methods is to empower future researchers with the requisite knowledge and tools, ensuring they are well-equipped to design and execute these studies effectively.

**Acknowledgements**
The authors would like to acknowledge and thank those who supported this work, including the Training in Implementation Science Research Methods T32HL155000 grant that supported Melody Owen, and the UG3HL154297 grant from the National Heart, Lung, and Blood Institute that supported Abel Kho, J.D. Smith, and Paris Davis. Support for J.D. Smith was also provided by the Dissemination and Implementation Science Core (DISC) of the Utah Clinical and Translational Science Institute (UM1 TR004409).

**Conflict of Interest**
The authors declare no potential conflict of interests.

**Tables and Figures**

*Table 1. Summary of Implementation Outcomes defined in RE-AIM Model[21]*

| Implementation Outcome | Definition | Example in Literature |
|---|---|---|
| **Reach** | Proportion of individuals that received the intervention among all individuals eligible to receive the intervention, measured at the individual level | Galaviz et al. 2017 – Reach was measured as the participation rate for patients who took part in the clinical intervention[42] |
| **Efficacy** | The degree to which the intervention is successful when implemented as intended, measured at the individual level | Houston et al. 2015 – Efficacy was measured as the 6-month cessation rate for smoking[43] |
| **Adoption** | Proportion of clusters that adopt/up-take the intervention, measured at the cluster level | Galaviz et al. 2017 – Adoption was measured as the rate of physicians who took the implementation strategy (training course) across the clusters [42] |
| **Implementation** | The degree to which the intervention is implemented as intended, measured at the cluster level | Houston et al. 2015 – Implementation was measured by the number of smokers referred and registered to program intervention[43] |
| **Maintenance** | The degree to which a program is continued and sustained over time, measured at both the individual and cluster level | Galaviz et al. 2017 – Maintenance was analyzed by collecting a follow-up questionnaire from patients 6 months after receiving clinical intervention[42] |

*Table 2. Description of Input Parameters Motivated by CIRCL[44,45] (2 Co-Primary Outcomes)*

| Input Parameter | Description | Statistical Notation | Value |
| --- | --- | --- | --- |
| Number of clusters | Number of clinics in each treatment arm | $K$ | 15 |
| Cluster Size | Number of patients in each clinic | $m$ | 300 |
| Statistical Power | Probability of detecting a true effect under the alternative hypothesis | $\pi$ | 80% |
| Overall (or family-wise) False Positive Rate | Probability of one or more Type I error(s) | $\alpha$ | 0.05 |
| Effect for $Y_1$ | Estimated intervention effect on hypertension control, in percentage point increase | $\beta_1^*$ | 10% |
| Effect for $Y_2$ | Estimated Intervention effect on reach, in percentage point increase | $\beta_2^*$ | 10% |
| Endpoint-specific ICC for $Y_1$ | Correlation for hypertension control ($Y_1$) for two different individuals in the same clinic | $\rho_0^{(1)} = Corr(Y_{1,ikj}, Y_{1,ikj'})$ | 0.025 |
| Endpoint-specific ICC for $Y_2$ | Correlation for reach ($Y_2$) for two different individuals in the same clinic | $\rho_0^{(2)} = Corr(Y_{2,ikj}, Y_{2,ikj'})$ | 0.025 |
| Inter-subject between-endpoint ICC | Correlation between hypertension control ($Y_1$) and reach ($Y_2$) for two *different* individuals in the same cluster | $Corr(Y_{1,ikj}, Y_{2,ikj'}) = \rho_1^{(1,2)}$ | 0.01 |
| Intra-subject between-endpoint ICC | Correlation between hypertension control ($Y_1$) and reach ($Y_2$) for the *same* individual | $Corr(Y_{1,ikj}, Y_{2,ikj}) = \rho_2^{(1,2)}$ | 0.05 |
| Total Variance of $Y_1$ | Total variance of the hypertension control outcome* | $Var(Y_1) = \sigma_1^2$ | 0.23 |
| Total Variance of $Y_2$ | Total variance of the reach outcome* | $Var(Y_2) = \sigma_2^2$ | 0.25 |

*Formula and calculation for total variance is available in Appendix B.1

*Table 3. Critical values for p-value adjustment methods*

| Method | | $\alpha$-level | Critical Value |
|---|---|---|---|
| Bonferonni | | $\alpha_{Bonf} = 0.025$ | 5.024 |
| Sidak | | $\alpha_{Sdk} = 0.0253$ | 5.002 |
| D/AP | $\rho_2^{(1,2)} = 0.01$ | $\alpha_{DAP} = 0.0255$ | 4.990 |
| | $\rho_2^{(1,2)} = 0.05$ | $\alpha_{DAP} = 0.0262$ | 4.943 |
| | $\rho_2^{(1,2)} = 0.1$ | $\alpha_{DAP} = 0.0271$ | 4.884 |

*Table 4. Summary of P-Value Adjustment Results for CIRCL hybrid type 2 study*

| P-Value Adjustment Method | Power ($\beta_1^*$) | Power ($\beta_2^*$) | K ($\beta_1^*$) | K ($\beta_2^*$) | m ($\beta_1^*$) | m ($\beta_2^*$) |
|---|---|---|---|---|---|---|
| Bonferroni | 87.62% | 84.55% | 12.35 | 13.43 | 104.76 | 148.58 |
| Sidak | 87.72% | 84.67% | 12.31 | 13.38 | 103.54 | 146.32 |
| D/AP | 87.99% | 84.98% | 12.21 | 13.27 | 100.36 | 140.53 |

*Table 5. Summary of five study design approaches for CIRCL hybrid 2 study*

| Method | $\chi^2$-distribution | | | F-distribution |
|---|---|---|---|---|
| | Power | K | m | Power |
| **1. P-Value Adjustments** | | | | |
|    a. Bonferroni | 84.55% | 14 | 149 | 80.45% |
|    b. Sidak | 84.67% | 14 | 147 | 80.61% |
|    c. D/AP | 84.98% | 14 | 141 | 81.02% |
| **2. Combined Outcomes** | 98.18% | 8 | 23 | 97.27% |
| **3. Single 1-DF Combined Test** | 98.11% | 8 | 23 | 97.29% |
| **4. Disjunctive 2-DF Test** | 96.01% | 9 | 34 | 93.63% |
| **5. Conjunctive IU Test*** | 91.43% | 11 | 74 | 89.92% |

*Conjunctive test uses MVN-distribution and t-distribution, respectively

*Figure 1. The significance level for a Type 2 Hybrid Design (Q=2) using the D/AP Method*

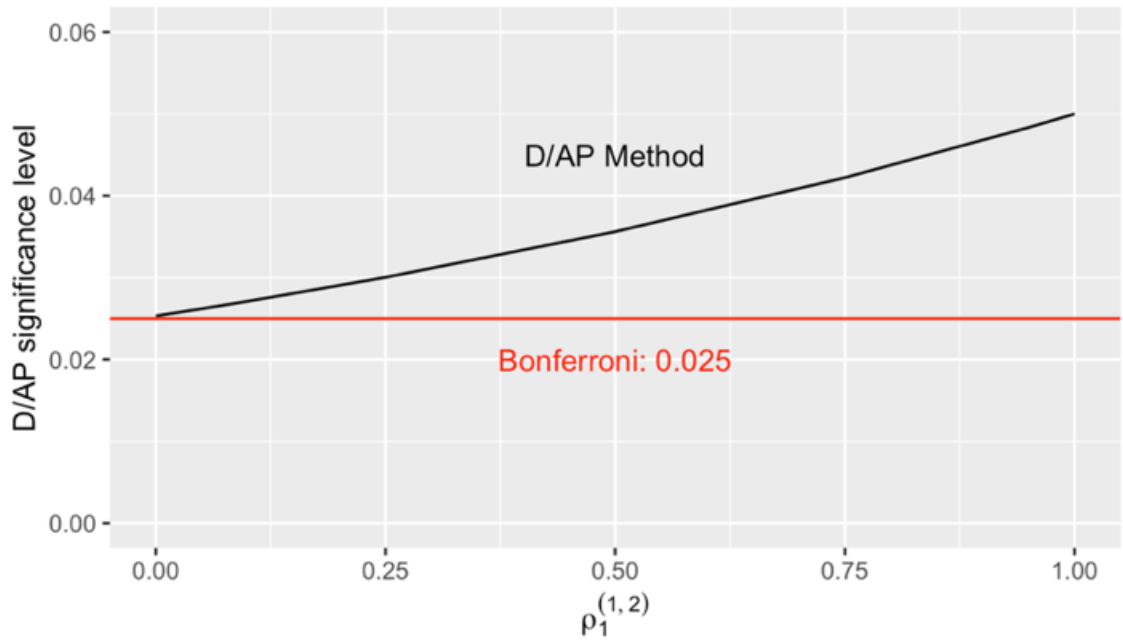

**Appendix A: Derivation of the Single 1-DF Combined Test for Cluster Randomized Trials**

The following shows an extension of the weighted combined test initially proposed by O'Brien et al. and extended by Pocock et al. for the case of clustered data.[34,35] A combined test statistic is useful when we have multiple endpoints that are related and are positively correlated, and the primary research goal is to test for the alternative hypothesis that at least one of the outcomes is significantly affected by the intervention.[34] The general case for multiple endpoints is given in an individual randomized controlled trial (RCT) scenario by Pocock et al., but here we focus on two outcomes.[34] For two primary endpoints, the test statistic given by Pocock et al. may be written as:

$$Z_{RCT}^2 = \left[\frac{\frac{Z_1 + Z_2}{2}}{\sqrt{\frac{1 + Corr(Z_1, Z_2)}{2}}}\right]^2 = \left[\frac{Z_1 + Z_2}{\sqrt{2(1 + Corr(Z_1, Z_2))}}\right]^2 \sim \chi_1^2 \text{ (under } H_0\text{)}$$

where $Z_1$ and $Z_2$ are the Z-test statistics for the first and second outcomes, respectively, and $Corr(Z_1, Z_2)$ is the correlation between the Z-test statistics of the first and second outcomes.[34] This follows from a Wald-type test statistic, which can be written as:

$$\frac{[Z_1 + Z_2 - 0]^2}{Var(Z_1 + Z_2)} = \frac{[Z_1 + Z_2]^2}{Var(Z_1) + Var(Z_2) + 2Cov(Z_1, Z_2)}$$

$$= \frac{[Z_1 + Z_2]^2}{Var(Z_1) + Var(Z_2) + 2\sqrt{Var(Z_1)Var(Z_2)}Corr(Z_1, Z_2)}$$

$$= \frac{[Z_1 + Z_2]^2}{2(1 + Corr(Z_1, Z_2))}$$

In an RCT, the correlation between $Z_1$ and $Z_2$ is straightforward and is usually expressed as $\rho$. This test statistic can be extended to the case of clustered data by deriving the expression for the correlation between $Z_1$ and $Z_2$, which requires consideration of the within and between-cluster correlation. The derivation for $Corr(Z_1, Z_2)$ in a clustered setting is shown in the following text. We begin with the general case of unequal treatment allocation, and then show the simplified form when the treatment allocation is equal.

Suppose we have a 2-arm cluster randomized trial with unequal treatment allocation. Let $r$ denote the treatment allocation ratio, and $i = 1,2$ denote the indices two treatment groups, where $K_1$ is the number of clusters assigned to the experimental group, and $K_2 = rK_1$ as the number of clusters assigned to the control group. The total number of clusters is $K_1 + K_2 = K_1 + rK_1 = (r + 1)K_1$, where clusters with index $k = 1, \ldots, K_1$ are in the treatment group and clusters with index $k = K_1 + 1, \ldots, (r + 1)K_1$ are in the control group. Let each cluster have $j = 1, \ldots, m$ individuals. Suppose we have two primary outcomes of interest, $Y_{q,ikj}$ (often referred to as simply $Y_q$), where $q = 1,2$. The intervention effect on a given outcome $Y_q$, denoted by $\beta_q^*$, is defined as the mean difference of $Y_q$ between the treatment group 1 and treatment group 2. Therefore, the point estimates of $\beta_q^*$ can be written as:

$$\hat{\beta}_q^* = \frac{1}{K_1 m}\sum_{k=1}^{K_1}\sum_{j=1}^{m} Y_{q,1kj} - \frac{1}{K_2 m}\sum_{k=K_1+1}^{K_1+K_2}\sum_{j=1}^{m} Y_{q,2kj}.$$

Assume that the following parameters are known; first, we have the total variance of outcome 1 and 2, $Var(Y_1) = \sigma_1^2$ and $Var(Y_2) = \sigma_2^2$, respectively. Next, we have the within-cluster correlation for outcome 1, also referred to as the ICC of outcome 1, $ICC(Y_1) =$ $\text{Corr}(Y_{1,ikj}, Y_{1,ikj'}) = \rho_0^{(1)}$ if $j \neq j'$, and $\text{Corr}(Y_{1,ikj}, Y_{1,ik'j'}) = 0$ if $k \neq k'$, for all $j, j'$. Similarly, for outcome 2, we have $ICC(Y_2) = \text{Corr}(Y_{2,ikj}, Y_{2,ikj'}) = \rho_0^{(2)}$ if $j \neq j'$, and $\text{Corr}(Y_{1,ikj}, Y_{1,ik'j'}) = 0$ if $k \neq k'$ for all $j, j'$. Then, we have the correlation of the two outcomes for two different individuals in the same cluster, namely $Corr(Y_{1,ikj}, Y_{2,ikj'}) = \rho_1^{(1,2)}$. Lastly, we have the correlation of the two outcomes for the same individual, namely $Corr(Y_{1,ikj}, Y_{2,ikj}) = \rho_2^{(1,2)}$.

Under this general setup with known parameters above and no assumption on the outcome model, the variance of these treatment effects can be written as:

$$\text{Var}(\hat{\beta}_1^*) = \text{Var}\left(\frac{1}{K_1 m}\sum_{k=1}^{K_1}\sum_{j=1}^{m} Y_{1,1kj}\right) + \text{Var}\left(\frac{1}{K_2 m}\sum_{k=K_1+1}^{K_1+K_2}\sum_{j=1}^{m} Y_{1,2kj}\right)$$

$$= \text{Var}\left(\frac{1}{K_1 m}\sum_{k=1}^{K_1}\sum_{j=1}^{m} Y_{1,1kj}\right) + \text{Var}\left(\frac{1}{rK_1 m}\sum_{k=K_1+1}^{(r+1)K_1}\sum_{j=1}^{m} Y_{1,2kj}\right)$$

$$= \frac{1}{(K_1 m)^2}\text{Var}\left(\sum_{k=1}^{K_1}\sum_{j=1}^{m} Y_{1,1kj}\right) + \frac{1}{(rK_1 m)^2}\text{Var}\left(\sum_{k=K_1+1}^{(r+1)K_1}\sum_{j=1}^{m} Y_{1,2kj}\right)$$

$$= \frac{1}{(K_1 m)^2}\left(\sum_{k=1}^{K_1}\sum_{j=1}^{m} \text{Var}(Y_{1,1kj})\right) + \frac{1}{(K_1 m)^2}\left(\sum_{k=1}^{K_1}\sum_{j\neq j'}^{m} \text{Cov}(Y_{1,1kj}, Y_{1,1kj'})\right)$$

$$+ \frac{1}{(K_1 m)^2}\left(\sum_{k\neq k'}^{K_1}\sum_{j,j'=1}^{m} \text{Cov}(Y_{1,1kj}, Y_{1,1k'j'})\right) + \frac{1}{(rK_1 m)^2}\left(\sum_{k=K_1+1}^{(r+1)K_1}\sum_{j=1}^{m} \text{Var}(Y_{1,2kj})\right)$$

$$+ \frac{1}{(rK_1 m)^2}\left(\sum_{k=K_1+1}^{(r+1)K_1}\sum_{j\neq j'}^{m} \text{Cov}(Y_{1,2kj}, Y_{1,2kj'})\right)$$

$$+ \frac{1}{(rK_1 m)^2}\left(\sum_{k=K_1+1\neq k'}^{(r+1)K_1}\sum_{j,j'=1}^{m} \text{Cov}(Y_{1,2kj}, Y_{1,2k'j'})\right)$$

(because $\text{Corr}(Y_{1,ikj}, Y_{1,ik'j'}) = 0$ if $k \neq k'$, the following equality hold)

$$= \frac{1}{(K_1 m)^2}\left(\sum_{k=1}^{K_1}\sum_{j=1}^{m} \text{Var}(Y_{1,1kj})\right) + \frac{1}{(K_1 m)^2}\left(\sum_{k=1}^{K_1}\sum_{j\neq j'}^{m} \text{Cov}(Y_{1,1kj}, Y_{1,1kj'})\right) + 0$$

$$+ \frac{1}{(rK_1 m)^2}\left(\sum_{k=K_1+1}^{(r+1)K_1}\sum_{j=1}^{m} \text{Var}(Y_{1,2kj})\right) + \frac{1}{(rK_1 m)^2}\left(\sum_{k=K_1+1}^{(r+1)K_1}\sum_{j\neq j'}^{m} \text{Cov}(Y_{1,2kj}, Y_{1,2kj'})\right)$$

$$+ 0$$

$$= \frac{\sigma_1^2}{K_1 m} + \frac{1}{(K_1 m)^2}\left(\sum_{k=1}^{K_1}\sum_{j=j'}^{m} \sigma_1^2 \rho_0^{(1)}\right) + \frac{\sigma_1^2}{rK_1 m} + \frac{1}{(rK_1 m)^2}\left(\sum_{k=K_1+1}^{(r+1)K_1}\sum_{j=j'}^{m} \sigma_1^2 \rho_0^{(1)}\right)$$

$$= \frac{\sigma_1^2}{K_1 m} + \frac{1}{(K_1 m)^2} K_1 m(m-1)\sigma_1^2 \rho_0^{(1)} + \frac{\sigma_1^2}{rK_1 m} + \frac{1}{(rK_1 m)^2} rK_1 m(m-1)\sigma_1^2 \rho_0^{(1)}$$

$$= \frac{\sigma_1^2}{K_1 m}\left(1 + (m-1)\rho_0^{(1)}\right) + \frac{\sigma_1^2}{rK_1 m}\left(1 + (m-1)\rho_0^{(1)}\right)$$

$$= \left(1 + \frac{1}{r}\right)\frac{\sigma_1^2}{K_1 m}\left(1 + (m-1)\rho_0^{(1)}\right)$$

The same follows for outcome $Y_2$, so the final variances of both outcome effects can be expressed as:

$$\text{Var}(\hat{\beta}_1^*) = \left(1 + \frac{1}{r}\right)\frac{\sigma_1^2}{K_1 m}\left(1 + (m-1)\rho_0^{(1)}\right); \quad \text{Var}(\hat{\beta}_2^*) = \left(1 + \frac{1}{r}\right)\frac{\sigma_2^2}{K_1 m}\left(1 + (m-1)\rho_0^{(2)}\right)$$

Then, using these derived variances of the estimated treatment effect, the z-test statistic for $Y_1$ and $Y_2$ can be written as:

$$Z_1 = \frac{\hat{\beta}_1^*}{\sqrt{\left(1 + \frac{1}{r}\right)\frac{\sigma_1^2}{K_1 m}\left(1 + (m-1)\rho_0^{(1)}\right)}}; \quad Z_2 = \frac{\hat{\beta}_2^*}{\sqrt{\left(1 + \frac{1}{r}\right)\frac{\sigma_2^2}{K_1 m}\left(1 + (m-1)\rho_0^{(2)}\right)}}$$

where under the null hypothesis, $Z_1 \sim \mathcal{N}(0,1)$ and $Z_2 \sim \mathcal{N}(0,1)$. Next, we derive the correlation between $Z_1$ and $Z_2$, namely $\text{Corr}(Z_1, Z_2)$. To do this, we first start with finding $\text{Cov}(\hat{\beta}_1^*, \hat{\beta}_2^*)$.

$$\text{Cov}(\hat{\beta}_1^*, \hat{\beta}_2^*) = \text{Cov}\left(\frac{1}{K_1 m}\sum_{k=1}^{K_1}\sum_{j=1}^{m} Y_{1,1kj} - \frac{1}{rK_1 m}\sum_{k=K_1+1}^{(r+1)K_1}\sum_{j=1}^{m} Y_{1,2kj}, \frac{1}{Km}\sum_{k=1}^{K_1}\sum_{j=1}^{m} Y_{2,1kj}\right.$$

$$\left. - \frac{1}{rK_1 m}\sum_{k=K_1+1}^{(r+1)K_1}\sum_{j=1}^{m} Y_{2,2kj}\right)$$

$$= \frac{1}{(K_1 m)^2}\text{Cov}\left(\sum_{k=1}^{K_1}\sum_{j=1}^{m} Y_{1,1kj} - \frac{1}{r}\sum_{k=K_1+1}^{(r+1)K_1}\sum_{j=1}^{m} Y_{1,2kj}, \sum_{k=1}^{K_1}\sum_{j=1}^{m} Y_{2,1kj} - \frac{1}{r}\sum_{k=K_1+1}^{(r+1)K_1}\sum_{j=1}^{m} Y_{2,2kj}\right)$$

$$= \frac{1}{(K_1 m)^2}\left[\text{Cov}\left(\sum_{k=1}^{K_1}\sum_{j=1}^{m} Y_{1,1kj}, \sum_{k=1}^{K_1}\sum_{j=1}^{m} Y_{2,1kj}\right) - \text{Cov}\left(\sum_{k=1}^{K_1}\sum_{j=1}^{m} Y_{1,1kj}, \frac{1}{r}\sum_{k=K_1+1}^{(r+1)K_1}\sum_{j=1}^{m} Y_{2,2kj}\right)\right.$$

$$- \text{Cov}\left(\frac{1}{r}\sum_{k=K_1+1}^{(r+1)K_1}\sum_{j=1}^{m} Y_{1,2kj}, \sum_{k=1}^{K_1}\sum_{j=1}^{m} Y_{2,1kj}\right)$$

$$\left. + \text{Cov}\left(\frac{1}{r}\sum_{k=K_1+1}^{(r+1)K_1}\sum_{j=1}^{m} Y_{1,2kj}, \frac{1}{r}\sum_{k=K_1+1}^{(r+1)K_1}\sum_{j=1}^{m} Y_{2,2kj}\right)\right]$$

Taking each component of the expression above individually, we have:

$$\text{Cov}\left(\sum_{k=1}^{K_1}\sum_{j=1}^{m} Y_{1,1kj}, \sum_{k=1}^{K_1}\sum_{j=1}^{m} Y_{2,1kj}\right) = \sum_{k=1}^{K_1}\sum_{j=1}^{m}\text{Cov}\left(Y_{1,1kj}, \sum_{k=1}^{K_1}\sum_{j=1}^{m} Y_{2,1kj}\right)$$

$$= \sum_{k=1}^{K_1}\sum_{j=1}^{m}\left(\rho_2^{(1,2)}\sqrt{\sigma_1^2 \sigma_2^2} + (m-1)\rho_1^{(1,2)}\sqrt{\sigma_1^2 \sigma_2^2}\right)$$

$$= K_1 m\left(\rho_2^{(1,2)} + (m-1)\rho_1^{(1,2)}\right)\sqrt{\sigma_1^2 \sigma_2^2}$$

$$\text{Cov}\left(\sum_{k=1}^{K_1}\sum_{j=1}^{m} Y_{1,1kj}, \frac{1}{r}\sum_{k=K_1+1}^{(r+1)K_1}\sum_{j=1}^{m} Y_{2,2kj}\right) = 0$$

$$\operatorname{Cov}\left(\frac{1}{r}\sum_{k=K_1+1}^{(r+1)K_1}\sum_{j=1}^{m} Y_{1,2kj}, \sum_{k=1}^{K}\sum_{j=1}^{m} Y_{2,1kj}\right) = 0$$

$$\operatorname{Cov}\left(\frac{1}{r}\sum_{k=K_1+1}^{(r+1)K_1}\sum_{j=1}^{m} Y_{1,2kj}, \frac{1}{r}\sum_{k=K_1+1}^{(r+1)K_1}\sum_{j=1}^{m} Y_{2,2kj}\right) = \frac{1}{r^2}\sum_{k=K_1+1}^{(r+1)K_1}\sum_{j=1}^{m}\operatorname{Cov}\left(Y_{1,2kj}, \sum_{k=K_1+1}^{(r+1)K_1}\sum_{j=1}^{m} Y_{2,2kj}\right)$$

$$= \frac{1}{r^2}\sum_{k=K_1+1}^{(r+1)K_1}\sum_{j=1}^{m}\left(\rho_2^{(1,2)}\sqrt{\sigma_1^2\sigma_2^2} + (m-1)\rho_1^{(1,2)}\sqrt{\sigma_1^2\sigma_2^2}\right)$$

$$= \frac{1}{r^2}rK_1 m\left(\rho_2^{(1,2)} + (m-1)\rho_1^{(1,2)}\right)\sqrt{\sigma_1^2\sigma_2^2}$$

$$= \frac{K_1 m}{r}\left(\rho_2^{(1,2)} + (m-1)\rho_1^{(1,2)}\right)\sqrt{\sigma_1^2\sigma_2^2}$$

Then, this leaves us with a final expression of the covariance of the treatment effects, shown below.

$$\operatorname{Cov}(\hat{\beta}_1^*, \hat{\beta}_2^*) = \frac{1}{(K_1 m)^2}\left[K_1 m\left(\rho_2^{(1,2)} + (m-1)\rho_1^{(1,2)}\right)\sqrt{\sigma_1^2\sigma_2^2} + \frac{K_1 m}{r}\left(\rho_2^{(1,2)} + (m-1)\rho_1^{(1,2)}\right)\sqrt{\sigma_1^2\sigma_2^2}\right]$$

$$= \frac{1}{K_1 m}\left[\left(\rho_2^{(1,2)} + (m-1)\rho_1^{(1,2)}\right)\sqrt{\sigma_1^2\sigma_2^2} + \frac{1}{r}\left(\rho_2^{(1,2)} + (m-1)\rho_1^{(1,2)}\right)\sqrt{\sigma_1^2\sigma_2^2}\right]$$

$$= \left(1+\frac{1}{r}\right)\frac{1}{K_1 m}\left[\left(\rho_2^{(1,2)} + (m-1)\rho_1^{(1,2)}\right)\sqrt{\sigma_1^2\sigma_2^2}\right]$$

Using this, and recalling that $\operatorname{Var}(Z_1) = \operatorname{Var}(Z_2) = 1$, we can write $\operatorname{Corr}(Z_1, Z_2)$ as follows:

$$\operatorname{Corr}(Z_1, Z_2) = \frac{\operatorname{Cov}(Z_1, Z_2)}{\operatorname{Var}(Z_1) \times \operatorname{Var}(Z_1)} = \frac{\operatorname{Cov}(Z_1, Z_2)}{1 \times 1} = \operatorname{Cov}(Z_1, Z_2)$$

$$= \operatorname{Cov}\left(\frac{\hat{\beta}_1^*}{\sqrt{\left(1+\frac{1}{r}\right)\frac{\sigma_1^2}{K_1 m}\left(1+(m-1)\rho_0^{(1)}\right)}}, \frac{\hat{\beta}_2^*}{\sqrt{\left(1+\frac{1}{r}\right)\frac{\sigma_2^2}{K_1 m}\left(1+(m-1)\rho_0^{(2)}\right)}}\right)$$

$$= \frac{\operatorname{Cov}(\hat{\beta}_1^*, \hat{\beta}_2^*)}{\sqrt{\left(1+\frac{1}{r}\right)\frac{\sigma_1^2}{K_1 m}\left(1+(m-1)\rho_0^{(1)}\right)}\sqrt{\left(1+\frac{1}{r}\right)\frac{\sigma_2^2}{K_1 m}\left(1+(m-1)\rho_0^{(2)}\right)}}$$

$$= \frac{\left(1+\frac{1}{r}\right)\frac{1}{K_1 m}\left[\left(\rho_2^{(1,2)} + (m-1)\rho_1^{(1,2)}\right)\sqrt{\sigma_1^2\sigma_2^2}\right]}{\sqrt{\left(1+\frac{1}{r}\right)^2\frac{\sigma_1^2\sigma_2^2}{(K_1 m)^2}\left(1+(m-1)\rho_0^{(1)}\right)\left(1+(m-1)\rho_0^{(2)}\right)}}$$

$$= \frac{\left(1+\frac{1}{r}\right)\frac{\sigma_1\sigma_2}{K_1 m}\left(\rho_2^{(1,2)} + (m-1)\rho_1^{(1,2)}\right)}{\left(1+\frac{1}{r}\right)\frac{\sigma_1\sigma_2}{K_1 m}\sqrt{\left(1+(m-1)\rho_0^{(1)}\right)\left(1+(m-1)\rho_0^{(2)}\right)}}$$

$$= \frac{\left(\rho_2^{(1,2)} + (m-1)\rho_1^{(1,2)}\right)}{\sqrt{\left(1 + (m-1)\rho_0^{(1)}\right)\left(1 + (m-1)\rho_0^{(2)}\right)}}$$

Then, a valid test statistic for this 1-df combined test can be written as:

$$Z_{combined}^2 = \left[\frac{Z_1 + Z_2}{\sqrt{2(1 + \text{Corr}(Z_1, Z_2))}}\right]^2$$

$$= \frac{\left(\dfrac{\hat{\beta}_1^*}{\sqrt{\left(1 + \dfrac{1}{r}\right)\dfrac{\sigma_1^2}{K_1 m}\left(1 + (m-1)\rho_0^{(1)}\right)}} + \dfrac{\hat{\beta}_2^*}{\sqrt{\left(1 + \dfrac{1}{r}\right)\dfrac{\sigma_2^2}{K_1 m}\left(1 + (m-1)\rho_0^{(2)}\right)}}\right)^2}{2\left(1 + \dfrac{\left(\rho_2^{(1,2)} + (m-1)\rho_1^{(1,2)}\right)}{\sqrt{\left(1 + (m-1)\rho_0^{(1)}\right)\left(1 + (m-1)\rho_0^{(2)}\right)}}\right)}$$

Note that when the treatment allocation ratio is 1, i.e. $r = 1$ and $K_1 = K_2 = K$, the test statistic simplifies to the following:

$$Z_{combined}^2 = \frac{\left(\dfrac{\hat{\beta}_1^*}{\sqrt{\dfrac{2\sigma_1^2}{K_1 m}\left(1 + (m-1)\rho_0^{(1)}\right)}} + \dfrac{\hat{\beta}_2^*}{\sqrt{\dfrac{2\sigma_2^2}{K_1 m}\left(1 + (m-1)\rho_0^{(2)}\right)}}\right)^2}{2\left(1 + \dfrac{\left(\rho_2^{(1,2)} + (m-1)\rho_1^{(1,2)}\right)}{\sqrt{\left(1 + (m-1)\rho_0^{(1)}\right)\left(1 + (m-1)\rho_0^{(2)}\right)}}\right)}$$

Alternatively, following the notation from the disjunctive 2-DF test for two outcomes, since we denote $VIF_1 = 1 + (m-1)\rho_0^{(1)}$, $VIF_2 = 1 + (m-1)\rho_0^{(2)}$, and $VIF_{12} = \rho_2^{(1,2)} + (m-1)\rho_1^{(1,2)}$, the test statistic for this 1-DF combined test can also be written as:

$$Z_{combined}^2 = \left[\frac{Z_1 + Z_2}{\sqrt{2(1 + \text{Corr}(Z_1, Z_2))}}\right]^2 = \frac{(Z_1 + Z_2)^2}{2\left(1 + \dfrac{VIF_{12}}{\sqrt{VIF_1 VIF_2}}\right)}$$

When we compare this test statistic in a CRT setting to the setting of RCTs, we see that the correlation term changes to incorporate the four correlations, $\rho_0^{(1)}$, $\rho_0^{(2)}$, $\rho_1^{(1,2)}$, and $\rho_2^{(1,2)}$, that are present in the case of clustered data.

## Appendix B: Equation Derivations

### B.1 Derivation of the Total Variance from the Within-Cluster Variance and ICC

Typically at the study design stage, investigators are able to provide a within-cluster cross-sectional variance of outcomes, since they usually can access to this quantity at, say, a single clinic, while study design formulas require the total variances of $Y_1$ and $Y_2$, denoted as $Var(Y_1) = \sigma_1^2$ and $Var(Y_2) = \sigma_2^2$. Assuming a linear outcome model in the identity link for each of the $Q = 2$ outcomes following model $Y_{q,ikj} = \mu_{q,i} + b_{q,ik} + e_{q,ikj}$, where $Var(b_{q,ik}) = \sigma_{q,B}^2$ is the between-cluster component of variance for outcome $q$, $Var(e_{q,ikj}) = \sigma_{q,W}^2$ is the within-cluster component of variance for outcome $q$, and the total variances of $Y_1$ and $Y_2$ are $Var(Y_1) = \sigma_1^2 = \sigma_{1,B}^2 + \sigma_{1,W}^2$ and $Var(Y_2) = \sigma_2^2 = \sigma_{2,B}^2 + \sigma_{2,W}^2$, respectively. For binary data, $\sigma_{q,W}^2 = p_q(1 - p_q)$, where $p_q$ is the proportion of individuals in a single cluster with $Y_q = 1$ (in CIRCL-Chicago, $p_1 = 0.66$ and $p_2 = 0.60$). As given in Table 2, let $\rho_0^{(1)} = 0.025$, $\rho_0^{(2)} = 0.025$. Then, $\sigma_{1,W}^2 = 0.66(1 - 0.66) = 0.22$ and $\sigma_{2,W}^2 = 0.60(1 - 0.60) = 0.24$. Now, we want to calculate the between-cluster component of variance for each outcome, namely $\sigma_{1,B}^2$ and $\sigma_{2,B}^2$, which can be accomplished by using the equation $\sigma_{q,B}^2 = \frac{\rho_q^{(1)} \sigma_{q,W}^2}{1-\rho_q^{(1)}}$. For the CIRCL-Chicago study, we obtain

$$\sigma_{1,B}^2 = \frac{\rho_0^{(1)} \sigma_{1,W}^2}{1-\rho_0^{(1)}} = \frac{0.025(0.22)}{1-0.025} = 0.0056, \text{ and } \sigma_{2,B}^2 = \frac{\rho_0^{(2)} \sigma_{2,W}^2}{1-\rho_0^{(2)}} = \frac{0.025(0.24)}{1-0.025} = 0.0062,$$

leading to the total outcome variances as $Var(Y_1) = \sigma_{1,B}^2 + \sigma_{1,W}^2 = 0.0056 + 0.22 = 0.2256 \approx 0.23$, and $Var(Y_2) = \sigma_{2,B}^2 + \sigma_{2,W}^2 = 0.0062 + 0.24 = 0.2462 \approx 0.25$.

### B.2 Derivation of the Endpoint Specific ICC for the Combined Outcome Test

$$\rho_0^{(c)} = Corr(Y_{1,ikj} + Y_{2,ikj}, Y_{1,ikj'} + Y_{2,ikj'})$$

$$= \frac{Cov(Y_{1,ikj} + Y_{2,ikj}, Y_{1,ikj'} + Y_{2,ikj'})}{\sqrt{Var(Y_{1,ikj} + Y_{2,ikj})}\sqrt{Var(Y_{1,ikj'} + Y_{2,ikj'})}}$$

$$= \frac{Cov(Y_{1,ikj}, Y_{1,ikj'}) + Cov(Y_{1,ikj}, Y_{2,ikj'}) + Cov(Y_{2,ikj}, Y_{1,ikj'}) + Cov(Y_{2,ikj}, Y_{2,ikj'})}{\sqrt{Var(Y_{1,ikj}) + Var(Y_{2,ikj}) + 2Cov(Y_{1,ikj}, Y_{2,ikj})}\sqrt{Var(Y_{1,ikj'}) + Var(Y_{2,ikj'}) + 2Cov(Y_{1,ikj'}, Y_{2,ikj'})}}$$

$$= \frac{\rho_0^{(1)} \sigma_1 \sigma_1 + \rho_1^{(1,2)} \sigma_1 \sigma_2 + \rho_1^{(1,2)} \sigma_1 \sigma_2 + \rho_0^{(2)} \sigma_2 \sigma_2}{\sqrt{Var(Y_{1,ikj}) + Var(Y_{2,ikj}) + 2Cov(Y_{1,ikj}, Y_{2,ikj})}\sqrt{Var(Y_{1,ikj'}) + Var(Y_{2,ikj'}) + 2Cov(Y_{1,ikj'}, Y_{2,ikj'})}}$$

$$= \frac{\rho_0^{(1)} \sigma_1^2 + \rho_0^{(2)} \sigma_2^2 + 2\rho_1^{(1,2)} \sigma_1 \sigma_2}{\sqrt{\sigma_1^2 + \sigma_2^2 + 2\rho_2^{(1,2)} \sigma_1 \sigma_2}\sqrt{\sigma_1^2 + \sigma_2^2 + 2\rho_2^{(1,2)} \sigma_1 \sigma_2}} = \frac{\rho_0^{(1)} \sigma_1^2 + \rho_0^{(2)} \sigma_2^2 + 2\rho_1^{(1,2)} \sigma_1 \sigma_2}{\sigma_1^2 + \sigma_2^2 + 2\rho_2^{(1,2)} \sigma_1 \sigma_2}$$

### B.3 Derivation of the Total Combined Outcome Variance

Using principles of probability, we can derive an expression of the total variance of the combined outcome, $Y_1 + Y_2$, as follows:

$$\sigma_c^2 = \sigma_1^2 + \sigma_2^2 + 2\,Cov(Y_{1,ikj}, Y_{2,ikj})$$

$$= \sigma_1^2 + \sigma_2^2 + 2(\sigma_1)(\sigma_2)\,\text{Corr}(Y_{1,ikj}, Y_{2,ikj})$$
$$= \sigma_1^2 + \sigma_2^2 + 2\rho_2^{(1,2)}\sigma_1\sigma_2$$

### B.4 Derivation of the Sidak Adjusted Significance Level[29]

$$P(\text{No Type I error on one test}) = 1 - \alpha_{Sdk}$$
$$P(\text{No Type I error on Q tests}) = (1 - \alpha_{Sdk})^Q$$
$$P(\text{At least one Type I error on Q tests}) = 1 - (1 - \alpha_{Sdk})^Q = \alpha$$
$$\alpha_{Sdk} = 1 - (1-\alpha)^{\frac{1}{Q}}$$
$$\alpha_{Sdk} = 1 - \sqrt{(1-\alpha)} \quad \text{for } Q = 2$$

### B.5 Proof of relationship between P-Value adjustment methods

Here, we prove that $\alpha^{Bonferroni} < \alpha^{Sidak} < \alpha^{DAP}$ for the general case, where $Q$ is the number of outcomes and can be greater than or equal to 2.

First, we will prove that $\alpha^{Bonferroni} < \alpha^{Sidak}$. Recall that $\alpha^{Bonferroni} = \frac{\alpha}{Q}$ and $\alpha^{Sidak} = 1 - (1-\alpha)^{1/Q}$. Now, Bernoulli's inequality states that $(1+x)^r < 1 + rx$ for $x > -1$ and $0 < r < 1$. Let $x = -\alpha$ and $r = 1/Q$. Then, it clearly follows that
$$(1-\alpha)^{1/Q} < 1 - \frac{\alpha}{Q}$$
and so we conclude that $\alpha^{Bonferroni} < \alpha^{Sidak}$.

Now, we will prove that $\alpha^{Sidak} < \alpha^{DAP}$. Recall that $\alpha^{Sidak} = 1 - (1-\alpha)^{1/Q}$ and $\alpha_q^{DAP} = 1 - (1-\alpha)^{\frac{1}{M_q}}$. Notice that because $M_q = Q^{1-\rho_2^{(1,2)}} < Q$, we have that $\frac{1}{M_q} > \frac{1}{Q}$. Then it follows that $(1-\alpha)^{\frac{1}{M_q}} < (1-\alpha)^{\frac{1}{Q}}$ as $0 < 1-\alpha < 1$, which further suggests $1 - (1-\alpha)^{\frac{1}{M_q}} > 1 - (1-\alpha)^{\frac{1}{Q}}$. This proves that $\alpha^{Sidak} < \alpha^{DAP}$, and we conclude that $\alpha^{Bonferroni} < \alpha^{Sidak} < \alpha^{DAP}$.

### B.6 Simplification of the noncentrality parameter for disjunctive 2-DF test

Recall the following equations from Yang et al., which says that under equal treatment allocation, the noncentrality parameter is $2K(L\beta^*)^T \left(L\,\Omega_{\beta^*}L^T\right)^{-1}L\beta^*$, where $\Omega_{\beta^*}$ is the variance-covariance matrix for the treatment effect estimate $\beta^*$, and

$$\omega_1^2 = \frac{4\sigma_1^2\left[1 + (m-1)\rho_0^{(1)}\right]}{m} = \frac{4\sigma_1^2\,VIF_1}{m}; \quad \omega_2^2 = \frac{4\sigma_2^2\left[1 + (m-1)\rho_0^{(2)}\right]}{m} = \frac{4\sigma_2^2\,VIF_2}{m};$$

$$\omega_{1,2} = \frac{4\sigma_1\sigma_2\left[\rho_2^{(1,2)} + (m-1)\rho_1^{(1,2)}\right]}{m} = \frac{4\sigma_1\sigma_2\,VIF_{12}}{m}$$

where $\omega_1^2$ and $\omega_2^2$ are the elements on the main diagonal of the 2 × 2 variance-covariance matrix, and $\omega_{qq'}$ is the element of the off-diagonal.[36] Then, for two co-primary outcomes and equal treatment allocation, the contrast matrix is $L = \begin{bmatrix} 1 & 0 \\ 0 & 1 \end{bmatrix}$, and we can simplify the equation as follows:

$$\lambda = 2K(L\beta^*)^T \left(L\,\Omega_{\beta^*}L^T\right)^{-1}L\beta^*$$

$$= 2K\left(\begin{bmatrix}1 & 0\\0 & 1\end{bmatrix}\begin{bmatrix}\beta_1^*\\\beta_2^*\end{bmatrix}\right)^T\left(\begin{bmatrix}1 & 0\\0 & 1\end{bmatrix}\begin{bmatrix}\omega_1^2 & \omega_{1,2}\\\omega_{1,2} & \omega_2^2\end{bmatrix}\begin{bmatrix}1 & 0\\0 & 1\end{bmatrix}\right)^{-1}\begin{bmatrix}1 & 0\\0 & 1\end{bmatrix}\begin{bmatrix}\beta_1^*\\\beta_2^*\end{bmatrix}$$

$$= 2K\begin{bmatrix}\beta_1^*\\\beta_2^*\end{bmatrix}^T\begin{bmatrix}\omega_1^2 & \omega_{1,2}\\\omega_{1,2} & \omega_2^2\end{bmatrix}^{-1}\begin{bmatrix}\beta_1^*\\\beta_2^*\end{bmatrix}$$

$$= 2K\begin{bmatrix}\beta_1^* & \beta_2^*\end{bmatrix}\frac{1}{\omega_1^2\omega_2^2-(\omega_{1,2})^2}\begin{bmatrix}\omega_2^2 & -\omega_{1,2}\\-\omega_{1,2} & \omega_1^2\end{bmatrix}\begin{bmatrix}\beta_1^*\\\beta_2^*\end{bmatrix}$$

$$= \frac{Km[(\beta_1^*)^2\sigma_2^2 VIF_2 + (\beta_2^*)^2\sigma_1^2\ VIF_1 - 2\beta_1^*\beta_2^*\sigma_1\sigma_2 VIF_{12}]}{2\sigma_1^2\sigma_2^2[VIF_1 VIF_2 - VIF_{12}^2]}$$

**Appendix C: Relationship between Method 2 (Combined Outcomes Test) and Method 3 (Single 1-DF Combined Test) for Cluster-Randomized Trials**

Recall that under equal treatment allocation, the non-centrality parameters for each method can be written as:

$$\lambda^{(Method\ 2)} = \left(\frac{Km}{2}\right)\frac{(\beta_1^* + \beta_2^*)^2}{\sigma_c^2\left[1 + (m-1)\rho_0^{(c)}\right]}$$

$$\lambda^{(Method\ 3)} = \left(\frac{Km}{2}\right)\frac{\left(\frac{\beta_1^*}{\sqrt{2\sigma_1^2\left[1+(m-1)\rho_0^{(1)}\right]}} + \frac{\beta_2^*}{\sqrt{2\sigma_2^2\left[1+(m-1)\rho_0^{(2)}\right]}}\right)^2}{\left(1 + \frac{\left(\rho_2^{(1,2)} + (m-1)\rho_1^{(1,2)}\right)}{\sqrt{\left(1+(m-1)\rho_0^{(1)}\right)\left(1+(m-1)\rho_0^{(2)}\right)}}\right)}$$

Recall that $\sigma_c^2 = \sigma_1^2 + \sigma_2^2 + 2\rho_2^{(1,2)}\sigma_1\sigma_2$, and $\rho_0^{(c)} = \frac{\rho_0^{(1)}\sigma_1^2 + \rho_0^{(2)}\sigma_2^2 + 2\rho_1^{(1,2)}\sigma_1\sigma_2}{\sigma_1^2 + \sigma_2^2 + 2\rho_2^{(1,2)}\sigma_1\sigma_2}$ (as shown in Appendix B). Both Methods 2 and 3 weight each treatment effect differently to create a single test statistic. To understand the relationship between the methods, we will evaluate under which conditions under which the methods are mathematically equivalent. Setting the equations equal to each other and simplifying their expressions, we have:

$$\left(\frac{Km}{2}\right)\frac{(\beta_1^* + \beta_2^*)^2}{\sigma_c^2\left[1 + (m-1)\rho_0^{(c)}\right]} = \left(\frac{Km}{2}\right)\frac{\left(\frac{\beta_1^*}{\sqrt{2\sigma_1^2\left[1+(m-1)\rho_0^{(1)}\right]}} + \frac{\beta_2^*}{\sqrt{2\sigma_2^2\left[1+(m-1)\rho_0^{(2)}\right]}}\right)^2}{\left(1 + \frac{\left(\rho_2^{(1,2)} + (m-1)\rho_1^{(1,2)}\right)}{\sqrt{\left(1+(m-1)\rho_0^{(1)}\right)\left(1+(m-1)\rho_0^{(2)}\right)}}\right)}$$

$$\frac{\left(\frac{\beta_1^*}{\sigma_c} + \frac{\beta_2^*}{\sigma_c}\right)^2}{\left(1 + (m-1)\rho_0^{(c)}\right)}$$

$$= \frac{\left(\frac{\beta_1^*}{\sqrt{2\sigma_1^2\left[1+(m-1)\rho_0^{(1)}\right]}} + \frac{\beta_2^*}{\sqrt{2\sigma_2^2\left[1+(m-1)\rho_0^{(2)}\right]}}\right)^2}{\left(1 + \frac{\left(\rho_2^{(1,2)} + (m-1)\rho_1^{(1,2)}\right)}{\sqrt{\left(1+(m-1)\rho_0^{(1)}\right)\left(1+(m-1)\rho_0^{(2)}\right)}}\right)}$$

$$\frac{\left(\dfrac{\beta_1^*}{\sqrt{\sigma_1^2 + \sigma_2^2 + 2\rho_2^{(1,2)}\sigma_1\sigma_2}} + \dfrac{\beta_2^*}{\sqrt{\sigma_1^2 + \sigma_2^2 + 2\rho_2^{(1,2)}\sigma_1\sigma_2}}\right)^2}{\left(1 + (m-1)\dfrac{\rho_0^{(1)}\sigma_1^2 + \rho_0^{(2)}\sigma_2^2 + 2\rho_1^{(1,2)}\sigma_1\sigma_2}{\sigma_1^2 + \sigma_2^2 + 2\rho_2^{(1,2)}\sigma_1\sigma_2}\right)}$$

$$= \frac{\left(\dfrac{\beta_1^*}{\sqrt{2\sigma_1^2\left[1+(m-1)\rho_0^{(1)}\right]}} + \dfrac{\beta_2^*}{\sqrt{2\sigma_2^2\left[1+(m-1)\rho_0^{(2)}\right]}}\right)^2}{\left(1 + \dfrac{\left(\rho_2^{(1,2)} + (m-1)\rho_1^{(1,2)}\right)}{\sqrt{\left(1+(m-1)\rho_0^{(1)}\right)\left(1+(m-1)\rho_0^{(2)}\right)}}\right)}$$

$$\frac{\dfrac{\beta_1^*}{\sqrt{\sigma_1^2 + \sigma_2^2 + 2\rho_2^{(1,2)}\sigma_1\sigma_2}} + \dfrac{\beta_2^*}{\sqrt{\sigma_1^2 + \sigma_2^2 + 2\rho_2^{(1,2)}\sigma_1\sigma_2}}}{\sqrt{1 + (m-1)\dfrac{\rho_0^{(1)}\sigma_1^2 + \rho_0^{(2)}\sigma_2^2 + 2\rho_1^{(1,2)}\sigma_1\sigma_2}{\sigma_1^2 + \sigma_2^2 + 2\rho_2^{(1,2)}\sigma_1\sigma_2}}}$$

$$= \frac{\dfrac{\beta_1^*}{\sqrt{2\sigma_1^2\left[1+(m-1)\rho_0^{(1)}\right]}} + \dfrac{\beta_2^*}{\sqrt{2\sigma_2^2\left[1+(m-1)\rho_0^{(2)}\right]}}}{\sqrt{1 + \dfrac{\left(\rho_2^{(1,2)} + (m-1)\rho_1^{(1,2)}\right)}{\sqrt{\left(1+(m-1)\rho_0^{(1)}\right)\left(1+(m-1)\rho_0^{(2)}\right)}}}}$$

$$\frac{\beta_1^* + \beta_2^*}{\sqrt{\sigma_1^2 + \sigma_2^2 + \sigma_1^2\rho_0^{(1)}(m-1) + \sigma_2^2\rho_0^{(2)}(m-1) + 2\sigma_1\sigma_2\rho_1^{(1,2)}(m-1) + 2\sigma_1\sigma_2\rho_2^{(1,2)}}}$$

$$= \frac{\beta_1^*}{\sqrt{2\sigma_1^2 + 2\sigma_1^2\rho_0^{(1)}(m-1) + 2\sigma_1^2\rho_2^{(1,2)}\frac{\sqrt{1+(m-1)\rho_0^{(1)}}}{\sqrt{1+(m-1)\rho_0^{(2)}}} + 2\sigma_1^2\rho_1^{(1,2)}(m-1)\frac{\sqrt{1+(m-1)\rho_0^{(1)}}}{\sqrt{1+(m-1)\rho_0^{(2)}}}}}$$

$$+ \frac{\beta_2^*}{\sqrt{2\sigma_2^2 + 2\sigma_2^2\rho_0^{(2)}(m-1) + 2\sigma_2^2\rho_2^{(1,2)}\frac{\sqrt{1+(m-1)\rho_0^{(2)}}}{\sqrt{1+(m-1)\rho_0^{(1)}}} + 2\sigma_2^2\rho_1^{(1,2)}(m-1)\frac{\sqrt{1+(m-1)\rho_0^{(2)}}}{\sqrt{1+(m-1)\rho_0^{(1)}}}}}$$

Now assume that the outcome specific ICCs and outcome variances are the same, i.e. $\rho_0^{(1)} = \rho_0^{(2)}$ and $\sigma_1^2 = \sigma_2^2$. Rename the outcome variance for both outcomes as $\sigma^2 = \sigma_1^2 = \sigma_2^2$ and rename the outcome ICC for both outcomes as $\rho_0 = \rho_0^{(1)} = \rho_0^{(2)}$. Then, we have:

$$\frac{\beta_1^* + \beta_2^*}{\sqrt{\sigma^2 + \sigma^2 + \sigma^2\rho_0(m-1) + \sigma^2\rho_0(m-1) + 2\sigma^2\rho_1^{(1,2)}(m-1) + 2\sigma^2\rho_2^{(1,2)}}}$$

$$= \frac{\beta_1^*}{\sqrt{2\sigma^2 + 2\sigma^2\rho_0(m-1) + 2\sigma^2\rho_2^{(1,2)}\frac{\sqrt{1+(m-1)\rho_0}}{\sqrt{1+(m-1)\rho_0}} + 2\sigma^2\rho_1^{(1,2)}(m-1)\frac{\sqrt{1+(m-1)\rho_0}}{\sqrt{1+(m-1)\rho_0}}}}$$

$$+ \frac{\beta_2^*}{\sqrt{2\sigma^2 + 2\sigma^2\rho_0(m-1) + 2\sigma^2\rho_2^{(1,2)}\frac{\sqrt{1+(m-1)\rho_0}}{\sqrt{1+(m-1)\rho_0}} + 2\sigma^2\rho_1^{(1,2)}(m-1)\frac{\sqrt{1+(m-1)\rho_0}}{\sqrt{1+(m-1)\rho_0}}}}$$

$$\frac{\beta_1^* + \beta_2^*}{\sqrt{2\sigma^2 + 2\sigma^2\rho_0(m-1) + 2\sigma^2\rho_1^{(1,2)}(m-1) + 2\sigma^2\rho_2^{(1,2)}}}$$

$$= \frac{\beta_1^*}{\sqrt{2\sigma^2 + 2\sigma^2\rho_0(m-1) + 2\sigma^2\rho_1^{(1,2)}(m-1) + 2\sigma^2\rho_2^{(1,2)}}}$$

$$+ \frac{\beta_2^*}{\sqrt{2\sigma^2 + 2\sigma^2\rho_0(m-1) + 2\sigma^2\rho_1^{(1,2)}(m-1) + 2\sigma^2\rho_2^{(1,2)}}}$$

$$\frac{\beta_1^* + \beta_2^*}{\sqrt{2\sigma^2 + 2\sigma^2\rho_0(m-1) + 2\sigma^2\rho_1^{(1,2)}(m-1) + 2\sigma^2\rho_2^{(1,2)}}}$$
$$= \frac{\beta_1^* + \beta_2^*}{\sqrt{2\sigma^2 + 2\sigma^2\rho_0(m-1) + 2\sigma^2\rho_1^{(1,2)}(m-1) + 2\sigma^2\rho_2^{(1,2)}}}$$

In conclusion, $\lambda^{(Method\ 3)} \to \lambda^{(Method\ 2)}$ as $\rho_0^{(1)} \to \rho_0^{(2)}$ and $\sigma_1^2 \to \sigma_2^2$. This is a sufficient condition, since $\rho_0^{(1)} = \rho_0^{(2)}$ and $\sigma_1^2 = \sigma_2^2$ results in $\lambda^{(Method\ 3)} = \lambda^{(Method\ 2)}$ regardless of the values of the remaining parameters. However, this is not necessarily a necessary condition, as it may be the case that $\lambda^{(Method\ 3)} = \lambda^{(Method\ 2)}$ for certain values when $\rho_0^{(1)} \neq \rho_0^{(2)}$ and/or $\sigma_1^2 \neq \sigma_2^2$.

# Appendix D: All design calculations for the CIRCL-Chicago Study

## D.1 P-value adjustment method calculations

$$\lambda^{(1)} = \frac{(\beta_1^*)^2}{2\frac{\sigma_1^2}{Km}[1+(m-1)\rho_0^{(1)}]} = \frac{(0.1)^2}{2\frac{0.23}{15(300)}[1+(300-1)0.025]} = 11.54$$

$$\lambda^{(2)} = \frac{(\beta_2^*)^2}{2\frac{\sigma_2^2}{Km}[1+(m-1)\rho_0^{(2)}]} = \frac{(0.1)^2}{2\frac{0.25}{15(300)}[1+(300-1)0.025]} = 10.62$$

$$\pi_{Bonf}^{(1)} = 1 - \Pr(\chi^2(1,\lambda^{(1)}) \leq 5.024) = 87.62\%$$
$$\pi_{Sidak}^{(1)} = 1 - \Pr(\chi^2(1,\lambda^{(1)}) \leq 5.002) = 87.72\%$$
$$\pi_{D/AP}^{(1)} = 1 - \Pr(\chi^2(1,\lambda^{(1)}) \leq 4.943) = 87.99\%$$
$$\pi_{Bonf}^{(2)} = 1 - \Pr(\chi^2(1,\lambda^{(2)}) \leq 5.024) = 84.55\%$$
$$\pi_{Sidak}^{(2)} = 1 - \Pr(\chi^2(1,\lambda^{(2)}) \leq 5.002) = 84.67\%$$
$$\pi_{D/AP}^{(2)} = 1 - \Pr(\chi^2(1,\lambda^{(2)}) \leq 4.943) = 84.98\%$$

$$\pi_{Bonf} = \min(\pi_{Bonf}^{(1)}, \pi_{Bonf}^{(2)}) = 84.55\%$$
$$\pi_{Sidak} = \min(\pi_{Sidak}^{(1)}, \pi_{Sidak}^{(2)}) = 84.67\%$$
$$\pi_{D/AP} = \min(\pi_{D/AP}^{(1)}, \pi_{D/AP}^{(2)}) = 84.98\%$$

$$K_{Bonf}^{(1)} = \frac{2(9.51)(0.23)[1+(300-1)0.025]}{300(0.1)^2} = 12.35 \approx 13$$

$$K_{Sidak}^{(1)} = \frac{2(9.47)(0.23)[1+(300-1)0.025]}{300(0.1)^2} = 12.31 \approx 13$$

$$K_{D/AP}^{(1)} = \frac{2(9.39)(0.23)[1+(300-1)0.025]}{300(0.1)^2} = 12.21 \approx 13$$

$$K_{Bonf}^{(2)} = \frac{2(9.51)(0.25)[1+(300-1)0.025]}{300(0.1)^2} = 13.43 \approx 14$$

$$K_{Sidak}^{(2)} = \frac{2(9.47)(0.25)[1+(300-1)0.025]}{300(0.1)^2} = 13.38 \approx 14$$

$$K_{D/AP}^{(2)} = \frac{2(9.39)(0.25)[1+(300-1)0.025]}{300(0.1)^2} = 13.27 \approx 14$$

$$K_{Bonf} = \max(K_{Bonf}^{(1)}, K_{Bonf}^{(2)}) = 14$$
$$K_{Sidak} = \max(K_{Sidak}^{(1)}, K_{Sidak}^{(2)}) = 14$$
$$K_{D/AP} = \max(K_{D/AP}^{(1)}, K_{D/AP}^{(2)}) = 14$$

$$m_{Bonf}^{(1)} = \frac{2(9.51)(0.23)(1-0.025)}{(0.1)^2 15 - 2(9.51)(0.23)(0.025)} = 104.76 \approx 105$$

$$m_{Sidak}^{(1)} = \frac{2(9.47)(0.23)(1-0.025)}{(0.1)^2 15 - 2(9.47)(0.23)(0.025)} = 103.54 \approx 104$$

$$m_{D/AP}^{(1)} = \frac{2(9.39)(0.23)(1-0.025)}{(0.1)^2 15 - 2(9.39)(0.23)(0.025)} = 100.36 \approx 101$$

$$m_{Bonf}^{(2)} = \frac{2(9.51)(0.25)(1-0.025)}{(0.1)^2 15 - 2(9.51)(0.25)(0.025)} = 148.58 \approx 149$$

$$m_{Sidak}^{(2)} = \frac{2(9.47)(0.25)(1-0.025)}{(0.1)^2 15 - 2(9.47)(0.25)(0.025)} = 146.32 \approx 147$$

$$m_{D/AP}^{(2)} = \frac{2(9.39)(0.25)(1-0.025)}{(0.1)^2 15 - 2(9.39)(0.25)(0.025)} = 140.53 \approx 141$$

$$m_{Bonf} = \max\left(m_{Bonf}^{(1)}, m_{Bonf}^{(2)}\right) = 149$$
$$m_{Sidak} = \max\left(m_{Sidak}^{(1)}, m_{Sidak}^{(2)}\right) = 147$$
$$m_{D/AP} = \max\left(m_{D/AP}^{(1)}, m_{D/AP}^{(2)}\right) = 141$$

D.2 Combined outcomes method calculations

$$\sigma_c^2 = \sigma_1^2 + \sigma_2^2 + 2\rho_2^{(1,2)}\sigma_1\sigma_2 = 0.23 + 0.25 + 2(0.05)\sqrt{(0.23)(0.25)} = 0.50$$

$$\rho_0^{(c)} = \frac{0.025(0.23) + 0.025(0.25) + 2(0.01)\sqrt{(0.23)(0.25)}}{0.23 + 0.25 + 2(0.05)\sqrt{(0.23)(0.25)}} = 0.03$$

$$\lambda = \frac{(0.2)^2}{2\frac{0.50}{15(300)}[1 + (300-1)(0.03)]} = 16.42$$

$$\pi = 1 - Pr(\chi^2(1,\lambda) \leq 3.84) = 98.18\%$$

$$K = \frac{2(7.85)(0.50)[1 + (300-1)(0.03)]}{300(0.2)^2} = 7.17 \approx 8$$

$$m = \frac{2(7.85)(0.50)(1-0.03)}{(0.2)^2(15) - 2(7.85)(0.50)(0.03)} = 22.42 \approx 23$$

D.3 Single 1-DF combined test method calculations

$$\lambda = \left[ \frac{\sqrt{\frac{(0.1)^2}{\frac{2(0.23)}{15(300)}[1+(300-1)0.025]}} + \sqrt{\frac{(0.1)^2}{\frac{2(0.25)}{15(300)}[1+(300-1)0.025]}}}{\sqrt{2\left(1 + \frac{0.05+(300-1)(0.01)}{\sqrt{(1+(300-1)0.025)(1+(300-1)0.025)}}\right)}} \right]^2 = 16.30$$

$$\pi = Pr(\chi^2(1,\lambda) > 3.84) = 98.11\%$$

$$K = \frac{2(7.85)\left(1 + \frac{0.05+(300-1)(0.01)}{\sqrt{(1+(300-1)0.025)(1+(300-1)0.025)}}\right)}{\left[\sqrt{\frac{(0.1)^2}{\frac{2(0.23)}{300}[1+(300-1)0.025]}} + \sqrt{\frac{(0.1)^2}{\frac{2(0.25)}{300}[1+(300-1)0.025]}}\right]^2} = 7.22 \approx 8$$

$$7.85 = \left[ \frac{\sqrt{\frac{(0.1)^2}{\frac{2(0.23)}{15(m)}[1+(m-1)0.025]}} + \sqrt{\frac{(0.1)^2}{\frac{2(0.25)}{15(m)}[1+(m-1)0.025]}}}{\sqrt{2\left(1 + \frac{0.05+(300-1)(0.01)}{\sqrt{(1+(300-1)0.025)(1+(300-1)0.025)}}\right)}} \right]^2$$

$$m = 22.69 \approx 23$$

D.4 Disjunctive 2-DF test method calculations

$$VIF_1 = 1 + (m-1)\rho_0^{(1)} = 1 + (300-1)(0.025) = 8.48$$
$$VIF_2 = 1 + (m-1)\rho_0^{(2)} = 1 + (300-1)(0.025) = 8.48$$
$$VIF_{12} = \rho_2^{(1,2)} + (m-1)\rho_1^{(1,2)} = 0.05 + (300-1)(0.01) = 3.04$$

$$\lambda = \left[ \frac{(15)(300)[(0.1)^2(0.25)(8.48) - 2(0.1)(0.1)\sqrt{0.23}\sqrt{0.25}(3.04) + (0.1)^2(0.23)(8.48)]}{2(0.23)(0.25)[(8.48)(8.48) - (3.04)^2]} \right]$$
$$= 16.32$$

$$\pi = 1 - \int_0^{5.99} f(x; 2, \lambda = 16.32) dx = 96.01\%$$

$$K = \left[\frac{2(9.63)(0.23)(0.25)[(8.48)(8.48) - (3.04)^2]}{(300)[(0.1)^2(0.25)(8.48) - 2(0.1)(0.1)\sqrt{0.23}\sqrt{0.25}(3.04) + (0.1)^2(0.23)(8.48)]]}\right]$$
$$= 8.86 \approx 9$$

$$9.63 = \left[\frac{15(m)[(0.1)^2(0.25)(VIF_2) - 2(0.1)(0.1)\sqrt{0.23}\sqrt{0.25}(VIF_{12}) + (0.1)^2(0.23)(8.48)]}{2(0.23)(0.25)[(VIF_1)(VIF_2) - (VIF_{12})^2]}\right]$$
$$m = 33.84 \approx 34$$

D.5 Conjunctive test method calculations

$$[\zeta_1, \zeta_2]^T = \left[\frac{(0.1)\sqrt{2(15)}}{\sqrt{\frac{4(0.23)(8.475)}{300}}}, \frac{(0.1)\sqrt{2(15)}}{\sqrt{\frac{4(0.25)(8.475)}{300}}}\right]^T = [3.40, 3.26]^T$$

$$c_1 = c_2 = t_\alpha(2 \times 15 - 4) = 1.71$$

Software is used to calculate the quantities: $\pi = 89.92\%$; $K = 12$; $m = 86$.